\documentclass[journal,onecolumn,12pt,draftcls,onecolumn]{IEEEtran}

\usepackage{epsfig,latexsym}
\usepackage{float}
\usepackage{indentfirst}
\usepackage{amsmath}
\usepackage{amssymb}
\usepackage{times}
\usepackage{subfigure}
\usepackage{psfrag}
\usepackage{cite}
\usepackage{color}
\definecolor{white}{rgb}{0.796,0.948,0.816}
\definecolor{black}{rgb}{0.00,0.00,0.00}

\def\diag{\textrm{diag}}

\newtheorem{theorem}{Theorem}
\newtheorem{corollary}{Corollary}

\begin{document}
\title{ {\color{black} Asymptotic Performance Analysis of GSVD-NOMA Systems with a Large-Scale Antenna Array} \thanks{}}

\vspace{-1em}
\author{Zhuo Chen,  Zhiguo Ding, \IEEEmembership{Senior Member, IEEE}, Xuchu Dai, and Robert Schober, \IEEEmembership{Fellow, IEEE}
\thanks{Zhuo Chen  and Xuchu Dai are with Key Lab of Wireless-Optical Commun., Chinese Acad. of Sciences, Sch. Info Science \& Tech., Univ. Science \& Tech. China, Hefei, Anhui, 230027, P.R.China (E-mail: cz1992@mail.ustc.edu.cn, daixc@ustc.edu.cn).

Zhiguo Ding is with the School of  Computer and Communications, Lancaster University, LA1 4YW, UK (E-mail: z.ding@lancaster.ac.uk).

Robert Schober is with the Institute for Digital Communications,
University Erlangen-N\"urnberg, Cauerstrasse 7, D-91058 Erlangen, Germany
(Email:  robert.schober@fau.de).
}}
\maketitle

{\color{black}}
\vspace{-4em}
\begin{abstract}
This paper considers  a multiple-input multiple-output (MIMO) downlink communication scenario with one base
station and two users, where each user is equipped with $m$
antennas and the  base
station is  equipped with $n$
antennas.
To efficiently exploit the spectrum resources, we propose
a transmission protocol which combines  generalized singular value decomposition (GSVD) and
non-orthogonal multiple
access (NOMA).
The average data rates achieved by
the two users are adopted as performance metrics for evaluation of
the proposed GSVD-NOMA scheme.
In particular,
we first characterize  {\color{black}the limiting distribution} of
the squared generalized singular values
of the two users' channel matrices
for the  asymptotic case where the numbers of
transmit and receive antennas approach infinity.
Then, we calculate
the normalized average individual rates of the users
in the considered asymptotic regime.
{\color{black}
Furthermore,
we extend
the proposed GSVD-NOMA scheme
to the MIMO
downlink communication scenario with
more than two users by
using a  hybrid multiple access (MA) approach,
where the base station first
divides
the users into different groups, then the proposed GSVD-NOMA scheme is
implemented within each group,
and different groups are allocated with
orthogonal bandwidth resources.}
Finally, numerical results are provided to validate
the effectiveness of the proposed GSVD-NOMA protocol, and the accuracy of the developed
analytical results.
\end{abstract}


\section{Introduction}
{\color{black}
Non-orthogonal multiple access (NOMA) is an effective approach to improve the spectral efficiency of wireless networks
and has been recognized as
a promising candidate for 5G multiple
access (MA)\cite{saito2013non}.
The key idea of NOMA is
to serve multiple users
at the same frequency resource blocks and at the same time.
In general,
NOMA can be implemented in two ways:
first, by using single-carrier NOMA \cite{ding2017application, xu2017optimal},
where
the principle of NOMA, spectrum sharing, is implemented on
one resource block, such as one subcarrier,
and, second, by
using multi-carrier NOMA \cite{sun2017optimal}, where
the principle of NOMA is jointly implemented across multiple orthogonal resource blocks.
In the past two years,  NOMA has been widely used and investigated
due to its superior
spectral efficiency and flexibility compared to conventional orthogonal
MA (OMA) \cite{higuchi2013non, haci2015performance}.
In \cite{ding2014performance}, the performance of a downlink NOMA network
with randomly deployed users was investigated.
An uplink NOMA transmission
scheme was proposed in \cite{al2014uplink} and its performance was evaluated.
Moreover,
practical forms of  multi-carrier NOMA,
such as sparse code multiple access (SCMA)
and low-density spreading (LDS) NOMA,
were proposed in
\cite{nikopour2013sparse, razavi2012receiver},
which introduce
redundancy via coding/spreading among multiple subcarriers   to facilitate {\color{black}interference cancelation.}}

On the other hand,
multiple-input multiple-output
(MIMO) techniques have also been identified as a key
enabling technology for improving the 5G system throughput.
It is well known that
if perfect channel state information at the transmitter (CSIT) is available,
the capacity region of MIMO broadcast
channels can be achieved by using dirty paper coding (DPC) \cite{weingarten2006capacity}.
However, due to its prohibitive computational
complexity, it is difficult to implement DPC in practice
\cite{chen2016med}.
{\color{black}
Compared to DPC,
MIMO-NOMA schemes have a relatively low computational complexity
at the expense of a small loss in performance \cite{chen2016med}.}
Therefore,  MIMO-NOMA systems have attracted considerable research interest.
In \cite{sun2015ergodic}, the ergodic rate was maximized
for MIMO-NOMA systems having statistical CSIT only.
{\color{black}
In
\cite{choi2016power},
a layered transmission scheme was
applied to MIMO-NOMA systems
and
an optimal power allocation policy  was developed.}
In \cite{ding2016application}, a hybrid
MIMO-NOMA scheme was proposed, where
users were grouped into small-size clusters. NOMA was implemented
within each cluster, and MIMO detection was used
to cancel inter-cluster interference.
{\color{black}
In \cite{ding2016mimo},
a new MIMO-NOMA scheme
based on QR decomposition was proposed and power allocation policies
for this scheme were investigated.}
{\color{black} In \cite{shin2017coordinated},
coordinated beamforming
techniques were developed to enhance the performance of
MIMO-NOMA communications in the presence of inter-cell
interference.}
{\color{black}
In \cite{ali2017non},
NOMA was applied to downlink multiuser MIMO cellular
systems and a linear beamforming technique was proposed to
cancel  inter-cluster interference. A more detailed literature review on MIMO-NOMA
can be found in \cite{ding2017survey}.}

{\color{black}
The generalized singular value decomposition (GSVD) is an efficient tool to
decompose the MIMO-NOMA channel into  parallel single-input single-output (SISO) channels,  such that the NOMA principle  can be applied to each SISO channel individually}\footnote{ {\color{black}
The proposed GSVD-NOMA scheme can be viewed as
an extension of conventional MIMO processing for a single-user point-to-point link where
the singular value decomposition (SVD) is performed to
convert
the MIMO channel into parallel SISO channels.}
}. In \cite{ma2016general}, the
GSVD was applied
to MIMO-NOMA uplink and downlink transmission.
However, the authors in
\cite{ma2016general}  only consider the special case
where all nodes are equipped with the same number of antennas.
Also, in \cite{ma2016general}, the performance evaluation of the proposed GSVD-NOMA scheme relied on
computer simulations, and more insightful analytical results are missing.
Motivated by this, in this paper, we apply GSVD-NOMA to a general MIMO downlink communication scenario with one base
station and two users, where each user is equipped with $m$
antennas  and the  base
station is  equipped with $n$
antennas. {\color{black}Moreover,
the average data rates achieved by the two users are adopted as  performance metrics
for  evaluation of the proposed GSVD-NOMA scheme.}

The main contributions of this paper are summarized as
follows.

\begin{itemize}
\item{
We characterize
{\color{black}the limiting distribution}  of
the squared generalized singular values
of the two users' channel matrices
for the asymptotic case where the numbers of transmit and receive
antennas approach infinity, i.e., $n,m \to \infty$ with $\frac{m}{n}\to \eta$, where $\eta$ is
a constant denoting the ratio of the numbers of receive and transmit
antennas.
To the best of the authors' knowledge,
{\color{black}the limiting distribution}  of
the squared generalized singular values in the considered asymptotic regime,
has not been characterized before.
{\color{black}Compared to the eigenvalues and conventional singular values \cite{tulino2004foundations}, it is more challenging
to characterize the distribution  of the squared generalized singular values,
{\color{black}since they depend on the
  channel matrices of both users.}}
{\color{black}
 Long-term power normalization
  is applied
at the base station.
In order to investigate  the impact of  power normalization on the proposed GSVD-NOMA scheme,} {\color{black}
we study the properties of the GSVD decomposition matrix and characterize
the   long-term  power normalization factor.
}
}

\item{Furthermore, in order to evaluate the performance of the proposed GSVD-NOMA scheme,
we characterize
{\color{black}
the normalized average individual  rates}\footnote{ {\color{black}
Consider a linear memoryless vector channel of the form $\mathbf{y}=\mathbf{H}\mathbf{x}+\mathbf{n}$, where
$\mathbf{H}$ is the $m \times n$ channel matrix,
$\mathbf{x}$ is the $n$-dimensional input vector, $\mathbf{y}$
is the $m$-dimensional output vector, and the $m$-dimensional vector $\mathbf{n}$ models the additive
circularly symmetric Gaussian noise. The normalized input-output mutual
information of this channel is defined as $\frac{1}{m} I (  \mathbf{x}; \mathbf{y} ) $ \cite{tulino2004foundations}.
Similarly, in this paper, we define the normalized rate of a user as $\frac{1}{m} R_{d}$, where $R_{d}$ is the overall data rate of the user.}}
of  the two users
in the considered asymptotic regime.
{\color{black}
The  developed analytical results are easy to evaluate the performance of the proposed GSVD-NOMA scheme
and
can help avoid extensive computer simulation
in the considered asymptotic case where
the numbers
of transmit and receive antennas are large.
Also,
when the base station and the users have moderate numbers of  antennas (e.g. $m=2, n=5$),
the derived analytical results  still provide good approximations,
as indicated by the presented  simulation results.}
{\color{black}In addition, simulation results are  provided to  corroborate the improvement of
the normalized
average sum rate} of the proposed GSVD-NOMA scheme
 compared to  ``conventional'' OMA  and
QR-NOMA  in \cite{ding2016mimo}. {\color{black}
Moreover, a hybrid NOMA scheme
is proposed to
extend
the proposed GSVD-NOMA scheme
to the MIMO
downlink communication scenario with
more than two users, where the base station first
divides
the users into different groups, then the proposed GSVD-NOMA scheme is
implemented within each group,
and different groups are allocated with
orthogonal bandwidth resources.
Also,
numerical results
are provided to  demonstrate
the performance of this hybrid NOMA scheme.
}}
\end{itemize}

The rest of the paper is organized as follows.
In Section~\ref{system},
we introduce the system model considered in this paper.
In Section~\ref{prodef},
we present the proposed  GSVD-NOMA design which efficiently exploits the spectral resources.
{\color{black}
In  Section~\ref{perfom},
we derive  new analytical results for the  squared generalized singular values
and
develop  analytical expressions for the  normalized average individual rates of  the two users.
Numerical results are provided in Section~\ref{se5}, and Section~\ref{se6} concludes this paper.}

   \vspace{-1em}
\section{System model}
\label{system}
In this paper, we consider a
general MIMO downlink communication scenario with one base
station and two users, where each user is equipped with $m$
antennas and the  base
station is  equipped with $n$
antennas.
{\color{black} We assume block fading, i.e., the user channels are constant for the transmission
of one codeword and change independently from one codeword to the next.}
The $m \times n$ channel matrix
from the base station to the $i$-th user, $i \in \{1,2\} $,
is denoted by $\mathbf{G}_i$.
The considered composite channels are modeled as
$\mathbf{G}_i=\frac{1}{\sqrt{d_i^\tau}}\mathbf{H}_i$,
where  $\mathbf{H}_i \in \mathbb{C}^{m \times n}$ models  the small-scale fading,
{\color{black}  $\frac{1}{\sqrt{d_i^\tau}}$ models large-scale path loss, $\tau$ is the path
loss exponent,
and $d_i$
denotes the distance between the base station and the $i$-th user, $i \in \{1,2\} $.
The locations of the users affect the user channels
    via large-scale  path loss $\frac{1}{\sqrt{d_i^\tau}}$.}
All  small-scale fading coefficients are assumed to be
independent and identically distributed  Rayleigh with unit variance, i.e., $\mathbf{H}_i$ is an $m \times n$ matrix
whose elements are mutually independent and identically distributed (i.i.d.)
complex Gaussian random variables with zero mean and unit variance.
{\color{black}
Full CSIT is assumed to be available at the base station.}

Denote the $n \times n$  precoding matrix at the base station by  $\mathbf{P}_b$. Then,
the base station transmits an $n \times 1$ signal vector $\mathbf{x}$
to the two users, where $\mathbf{x}=\mathbf{P}_b \mathbf{s} $ and the $n \times 1$  vector $\mathbf{s}=[s_1,s_2, \cdots s_n]^T$
is the information bearing vector.
{\color{black} The
elements of  $\mathbf{s}$, $s_i$, $i \in \{1,\cdots,n\}$, are coded symbols representing the
messages intended for the two users and
are taken from  Gaussian codebooks, see Section~\ref{GSVDNOMA} for detailed description.}
The observations at the two users can be expressed as follows:
\begin{eqnarray}
\mathbf{y}_1=\frac{1}{\sqrt{d_1^\tau}} \mathbf{H}_1 \mathbf{x} +\mathbf{n}_1   \quad \text{and}  \quad \mathbf{y}_2=\frac{1}{\sqrt{d_2^\tau}} \mathbf{H}_2 \mathbf{x} +\mathbf{n}_2,
\end{eqnarray}
respectively, {\color{black} where $\mathbf{n}_i$, $i \in \{1,2\}$, denotes the Gaussian additive noise vector of user $i$,
whose elements are mutually independent and identically distributed (i.i.d.)
complex Gaussian random variables with zero mean and unit variance.} Define the detection matrices at user $1$ and user $2$  as {\color{black}
$\mathbf{D}_1 \in \mathbb{C}^{m \times m}$ and $\mathbf{D}_2 \in \mathbb{C}^{m \times m}$, respectively.} After  applying the detection matrices,
user $1$ and user $2$  observe the following signals:
\begin{eqnarray}
\label{rec2}
\mathbf{D}_1 \mathbf{y}_1= \frac{1}{\sqrt{d_1^\tau}} \mathbf{D}_1 \mathbf{H}_1 \mathbf{P}_b \mathbf{s}   +\mathbf{n}_1'
\quad \text{and}  \quad
\mathbf{D}_2  \mathbf{y}_2=  \frac{1}{\sqrt{d_2^\tau}} \mathbf{D}_2 \mathbf{H}_2 \mathbf{P}_b  \mathbf{s}  +\mathbf{n}_2',
\end{eqnarray}
respectively,
where $\mathbf{n}_1'=\mathbf{D}_1\mathbf{n}_1 $ and $\mathbf{n}_2'=\mathbf{D}_2\mathbf{n}_2 $.

   \vspace{-1em}
\section{Description of the Proposed GSVD-NOMA Scheme}
\label{prodef}

In the following, we propose a transmission protocol which combines  GSVD and NOMA.
To be more specific, we  first introduce
the GSVD
of the two $m \times n$ channel matrices $\mathbf{H}_1$ and $\mathbf{H}_2$.
{\color{black}
Please note that in this paper, we assume that
$\mathbf{H}_1$ and $\mathbf{H}_2$ are full rank\footnote{
{\color{black}Since the elements of $\mathbf{H}_1$ and $\mathbf{H}_2$
are i.i.d. complex Gaussian random variables, they are full rank with probability one
\cite{As09}.}}.}
Then, we explain the
designs of $\mathbf{P}_b$,  $\mathbf{D}_1$, $\mathbf{D}_2$, and   $\mathbf{s}$,
which are based on GSVD and  NOMA to efficiently  exploit the spectrum resources.
   \vspace{-1em}
\subsection{Definition of GSVD}
\label{GSVDdef}
We first introduce the GSVD
of the two $m \times n$ channel matrices $\mathbf{H}_1$ and $\mathbf{H}_2$
as follows \cite{zhaGSVD}:
\begin{eqnarray}
\label{GSVDFORM}
\mathbf{U}\mathbf{H}_1\mathbf{Q}=\mathbf{\Sigma}_1 \quad \text{and}  \quad \mathbf{V}\mathbf{H}_2\mathbf{Q}=\mathbf{\Sigma}_2,
\end{eqnarray}
where $\mathbf{U} \in \mathbb{C}^{m \times m}$ and $\mathbf{V} \in \mathbb{C}^{m \times m}$ are  two unitary matrices, $\mathbf{Q} \in \mathbb{C}^{n \times n}$ is a nonsingular matrix, and $\mathbf{\Sigma}_1 \in \mathbb{C}^{m \times n} $ and $\mathbf{\Sigma}_2 \in \mathbb{C}^{m \times n}$ are two
nonnegative diagonal matrices.
Moreover, $\mathbf{\Sigma}_1$ and $\mathbf{\Sigma}_2$ have the following forms depending on the choices of $m$ and $n$.
\begin{itemize}
\item{When $m \geq n$,
$\mathbf{\Sigma}_1$ and $\mathbf{\Sigma}_2$  can be expressed as follows:
\begin{eqnarray}
\label{mdayun}
\mathbf{\Sigma}_1= \left(\begin{array}{c}
\mathbf{S}_1\\
\mathbf{O}_{(m-n) \times n}
\end{array}\right)
\quad \text{and}  \quad \mathbf{\Sigma}_2=
\left(\begin{array}{c}
\mathbf{O}_{(m-n) \times n}\\
\mathbf{S}_2
\end{array}\right),
\end{eqnarray}
where
{\color{black}
$\mathbf{O}_{(m-n) \times n}$ denotes the zero matrix of size $(m-n) \times n$,
$\mathbf{S}_1= \diag(\alpha_1,\cdots,\alpha_n)$
and $\mathbf{S}_2=\diag(\beta_1,\cdots,\beta_n)$
are two $n \times n$ nonnegative
diagonal  matrices,
satisfying
$1\geq \alpha_1 \geq \alpha_2 \cdots \geq  \alpha_n \geq 0$
and
$\mathbf{S}_1^2 + \mathbf{S}_2^2 = \mathbf{I}_n$, {\color{black}
where $\mathbf{I}_n$ denotes the identity matrix of size $n$.}
Then, the squared generalized singular values are defined as $w_i^2=\alpha_i^2/\beta_i^2$, $i \in \{1,\cdots,n\}$.
}}

\item{When $m < n<2m$ ,   let us define  $r=n-m$ and {\color{black}$q=2m-n$}.
Then, $\mathbf{\Sigma}_1$ and $\mathbf{\Sigma}_2$  can be expressed as follows:
\begin{eqnarray}
\label{mmn1}
\mathbf{\Sigma}_1= \left(\begin{array}{ccc}
\mathbf{I}_r&\mathbf{O}_{r \times q} & \mathbf{O}_{r \times r}\\
\mathbf{O}_{q \times r}&\mathbf{S}_1&\mathbf{O}_{q \times r}
\end{array}\right)
\quad \text{and}  \quad \mathbf{\Sigma}_2=
\left(\begin{array}{ccc}
\mathbf{O}_{q \times r}&\mathbf{S}_2&\mathbf{O}_{q \times r}\\
\mathbf{O}_{r \times r}&\mathbf{O}_{r \times q}&\mathbf{I}_r
\end{array}\right),
\end{eqnarray}
where  $\mathbf{S}_1= \diag(\alpha_1,\cdots,\alpha_q)$
and $\mathbf{S}_2=\diag(\beta_1,\cdots,\beta_q)$
are two $q \times q$ nonnegative
diagonal  matrices, satisfying
$1\geq \alpha_1 \geq \alpha_2 \cdots \geq  \alpha_q \geq 0$
and
$\mathbf{S}_1^2 + \mathbf{S}_2^2 = \mathbf{I}_q$. {\color{black}Similarly, we define the squared generalized singular values as $w_i^2=\alpha_i^2/\beta_i^2$, $i \in \{1,\cdots,q\}$.}
}

\item{When $2m \leq n$,
$\mathbf{\Sigma}_1$ and $\mathbf{\Sigma}_2$  can be expressed as follows:
\begin{eqnarray}
\label{mmn2}
\mathbf{\Sigma}_1= \left(\begin{array}{cc}
\mathbf{I}_m & \mathbf{O}_{m \times (n-m)}
\end{array}\right)
\quad \text{and}  \quad \mathbf{\Sigma}_2=
\left(\begin{array}{cc}
 \mathbf{O}_{m \times (n-m)} & \mathbf{I}_m
\end{array}\right).
\end{eqnarray}
}
{\color{black}Note that the generalized singular values are independent of  $\mathbf{H}_1$ and $\mathbf{H}_2$ in this case.}
\end{itemize}

   \vspace{-1em}
\subsection{The proposed GSVD-NOMA design}
\label{GSVDNOMA}
As shown in \eqref{GSVDFORM}, the GSVD of the two
$m \times n$ matrices $\mathbf{H}_1$ and $\mathbf{H}_2$
can be expressed as $\mathbf{U}\mathbf{H}_1\mathbf{Q}=\mathbf{\Sigma}_1 \quad \text{and}  \quad \mathbf{V}\mathbf{H}_2\mathbf{Q}=\mathbf{\Sigma}_2$. Then,  the precoding matrix
$\mathbf{P}_b$
is given by   $\mathbf{P}_b=\mathbf{Q}  \sqrt{P}/ t$,
where  $P$ denotes the total transmission power of the  base station, and $t$ is a scalar for
power normalization.
{\color{black}
In this paper, for the sake of analytical tractability,
 long-term power normalization  is applied
at the base station, i.e.,  $t^2= \mathcal{E} \{ \text{trace} (\mathbf{Q}
\mathbf{s} \mathbf{s}^H \mathbf{Q}^H) \}$, where  $\mathbf{s}$ is the information bearing symbol vector, $\mathcal{E}\{ \cdot \}$ denotes mathematical expectation,
{\color{black} and $\text{trace}(\mathbf{Q}
\mathbf{s} \mathbf{s}^H \mathbf{Q}^H )$ denotes the trace of $\mathbf{Q}
\mathbf{s} \mathbf{s}^H \mathbf{Q}^H$}.}
The detection matrices $\mathbf{D}_1$ and $\mathbf{D}_2$ are chosen as
$\mathbf{D}_1=\mathbf{U}$ and $\mathbf{D}_2=\mathbf{V}$, respectively.
From \eqref{rec2}, we conclude that, with these choices,
user $1$ and user $2$ obtain the following signals:
{\color{black}
\begin{eqnarray}
\mathbf{U} \mathbf{y}_1=\frac{1}{\sqrt{d_1^\tau}} \mathbf{U} \mathbf{H}_1 \mathbf{P}_b  \mathbf{s}   +\mathbf{n}_1'=\frac{\sqrt{P}}{t\sqrt{d_1^\tau} }\mathbf{\Sigma}_1\mathbf{s} +\mathbf{n}_1',
\nonumber  \quad\text{and}
\end{eqnarray}
\begin{eqnarray}\mathbf{V} \mathbf{y}_2=
 \frac{1}{\sqrt{d_2^\tau}}  \mathbf{V} \mathbf{H}_2 \mathbf{P}_b \mathbf{s} +\mathbf{n}_2'=\frac{\sqrt{P}}{t\sqrt{d_2^\tau} } \mathbf{\Sigma}_2\mathbf{s} +\mathbf{n}_2',
\end{eqnarray}}where $\mathbf{n}_1'=\mathbf{U}\mathbf{n}_1 $ and $\mathbf{n}_2'=\mathbf{V}\mathbf{n}_2 $.
Next, we explain the design of $\mathbf{s}$ which is based on NOMA in order to fully exploit the available spectrum resources.

\subsubsection{The case when  $m \geq n$}
\label{m>n}
{\color{black}
As shown in Section~\ref{GSVDdef},
when $m \geq n$,
$\mathbf{\Sigma}_1$ and $\mathbf{\Sigma}_2$  can be expressed as in
\eqref{mdayun}.}
So, when $m \geq n$, after applying  detection matrices  $\mathbf{U}$ and  $\mathbf{V}$,
we obtain the following:
\begin{eqnarray}
\mathbf{U} \mathbf{y}_1=\frac{\sqrt{P}}{t\sqrt{d_1^\tau} }\left[ \begin{array}{c}
\alpha_1 s_1\\
\vdots\\
\alpha_n s_n \\
\mathbf{O}_{(m-n) \times 1}
\end{array} \right] +\mathbf{n}_1'
\quad \text{and}  \quad
\mathbf{V} \mathbf{y}_2=\frac{\sqrt{P}}{t\sqrt{d_2^\tau} }\left[ \begin{array}{c}
\mathbf{O}_{(m-n) \times 1}\\
\beta_1 s_1\\
\vdots\\
\beta_n s_n
\end{array} \right] +\mathbf{n}_2'.
\end{eqnarray}
Therefore, the use of the GSVD  converts
the downlink MIMO  channel into $n$ parallel SISO channels as follows:
\begin{eqnarray}
\label{para}
y_{1i}=\frac{\sqrt{P}}{t\sqrt{d_1^\tau} }
\alpha_i s_i +n_{1i}
\quad \text{and}  \quad
y_{2i}=\frac{\sqrt{P}}{t\sqrt{d_2^\tau} }
\beta_i s_i +n_{2i},  \quad i \in \{1,\cdots,n\},
\end{eqnarray}
{\color{black}
where $y_{1i}$ and $y_{2i}$ are the received signals at user 1 and user 2, respectively.}
As in \cite{ding2014performance}, we consider fixed-NOMA (F-NOMA) in this paper. To be more specific, let us define
two predefined power allocation coefficients $l_1$ and $l_2$
with $l_1>l_2$ and $l_1^2+l_2^2=1$.
For the $i$-th SISO channel,
when $\frac{\alpha_i}{\sqrt{d_1^\tau}} >\frac{\beta_i}{\sqrt{d_2^\tau}}$, i.e., user 1 has a stronger channel gain than user 2,
$s_i$ is designed as $s_i=l_2 s_{1i}+l_1 s_{2i}$, where $s_{1i}$
and $s_{2i}$ are the information bearing messages for user $1$ and user $2$, respectively.
On the other hand,
for the $i$-th SISO channel,
when $\frac{\alpha_i}{\sqrt{d_1^\tau}} \leq \frac{\beta_i}{\sqrt{d_2^\tau}}$, i.e., user 2 has a stronger channel gain than user 1,
$s_i$ is designed as $s_i=l_1 s_{1i}+l_2 s_{2i}$.
{\color{black}
Since Gaussian codebooks are
employed, the
input symbols $s_{1i}$ and $s_{2i}$ are independent  zero mean complex Gaussian random variables.
Moreover, as shown in \eqref{para}, after
applying the proposed GSVD-NOMA scheme,
the base station
decomposes the MIMO channels of the two users
into parallel SISO channels.
For simplicity, it is assumed that the
base station allocates the same power  to each
parallel SISO channel\footnote{ {\color{black}
More sophisticated, optimized
power allocation strategies
can further  increase the sum rate of the two users,
but make implementation and performance analysis
more complicated. Nevertheless, the development
of
optimal
power allocation strategies,
which maximize the sum rate and
ensure  user fairness at the same time, is an important direction of future research.}}, i.e.,
$\mathcal{E} \{  |s_{i}|^2 \}=\mathcal{E} \{  |s_{1i}|^2 \}=\mathcal{E} \{  |s_{2i}|^2 \}=1$, $i \in \{1,\cdots,n\}$.}
Recall that $w_i^2=\alpha_i^2/\beta_i^2$, $i \in \{1,\cdots,n\}$, are  the squared  generalized singular values.
Then, when  $\frac{\alpha_i}{\sqrt{d_1^\tau}} >\frac{\beta_i}{\sqrt{d_2^\tau}}$, i.e.,
$w_i^2 > \frac{d_1^\tau}{d_2^\tau}$,  SIC is performed at user $1$ to decode  $s_{1i}$ and
the information rates of $s_{1i}$ and $s_{2i}$ can be expressed as
\begin{eqnarray}
\label{m>n1>2R}
R_{1i}=\log \left(  1+\frac{P \alpha_i^2  l_2^2}{ t^2 d_1^\tau N_0} \right)
\quad \text{and}  \quad
R_{2i}=\log \left(  1+\frac{P \beta_i^2  l_1^2}{P \beta_i^2  l_2^2+ t^2 d_2^\tau N_0} \right),
\end{eqnarray}
respectively, where $N_0$ denotes the noise power. On the other hand, when
$\frac{\alpha_i}{\sqrt{d_1^\tau}} \leq \frac{\beta_i}{\sqrt{d_2^\tau}}$, i.e.,
$w_i^2 \leq \frac{d_1^\tau}{d_2^\tau}$,  SIC is performed at user $2$ to decode  $s_{2i}$ and
the information rates of $s_{1i}$ and $s_{2i}$ can be expressed as
\begin{eqnarray}
\label{m>n1<2R}
R_{1i}=\log \left(  1+\frac{P \alpha_i^2  l_1^2}{P \alpha_i^2  l_2^2+ t^2 d_1^\tau N_0} \right)
\quad \text{and}  \quad
R_{2i}=\log \left(  1+\frac{P \beta_i^2  l_2^2}{ t^2 d_2^\tau N_0} \right),
\end{eqnarray}
respectively.
{\color{black}
Note that \eqref{m>n1>2R} and  \eqref{m>n1<2R}
provide
the achievable instantaneous
rates of the users
for given user channel matrices
$\mathbf{H}_1$ and $\mathbf{H}_2$.}

\subsubsection{The case when   $m<n<2m$}
\label{m<n2m>n}
{\color{black}
As shown in Section~\ref{GSVDdef},
when  $m<n<2m$,
$\mathbf{\Sigma}_1$ and $\mathbf{\Sigma}_2$  can be expressed as in \eqref{mmn1}}.
So, when $m<n<2m$, after applying detection matrices  $\mathbf{U}$ and  $\mathbf{V}$,
user $1$ and user $2$ can observe the following signals:
\begin{eqnarray}
\label{papr}
\mathbf{U} \mathbf{y}_1=\frac{\sqrt{P}}{t\sqrt{d_1^\tau} }\left[ \begin{array}{c}
s_1\\
\vdots\\
s_r\\
\alpha_1 s_{r+1}\\
\vdots\\
\alpha_s s_{m}
\end{array} \right] +\mathbf{n}_1'
\quad \text{and}  \quad
\mathbf{V} \mathbf{y}_2=\frac{\sqrt{P}}{t\sqrt{d_2^\tau} }\left[ \begin{array}{c}
\beta_1 s_{r+1}\\
\vdots\\
\beta_s s_{m}\\
s_{m+1}\\
\vdots\\
s_n
\end{array} \right] +\mathbf{n}_2',
\end{eqnarray}
respectively.
{\color{black}
Again,
we assume that
$\mathcal{E} \{  |s_{i}|^2 \}=1$, $i \in \{1,\cdots,n\}$.}
For $i \in \{1,\cdots,r\}$,
$s_i$  is  observed by user $1$ only, and
{\color{black}the observations of $s_i$, $i \in \{1,\cdots,r\}$, at the two users
can be expressed as
\begin{eqnarray}
y_{1i}=\frac{\sqrt{P}}{t\sqrt{d_1^\tau} }
s_i +n_{1i}
\quad \text{and}  \quad
y_{2i}=0,
\end{eqnarray}
respectively.
}So, when $i \in \{1,\cdots,r\}$,
{\color{black}
we adopt the OMA transmission strategy, and} $s_i$ is designed as $s_i=s_{1i}$, i.e., $s_i$ only contains the
message for user $1$.
Therefore,
the corresponding information rate of $s_{1i}$, $i \in \{1,\cdots,r\}$,  can be expressed as
$
R_{1i}=\log \left(  1+\frac{P }{t^2 d_1^\tau N_0} \right).
$
When $i \in \{m+1,\cdots,n\}$, $s_i$  is observed by user $2$ only and
{\color{black}the observations of $s_i$, $i \in \{m+1,\cdots,n\}$, at the two users
can be expressed as
\begin{eqnarray}
y_{1i}=0
\quad \text{and}  \quad
y_{2i}=\frac{\sqrt{P}}{t\sqrt{d_2^\tau} }
s_i +n_{2i},
\end{eqnarray}
respectively.}
Hence,  when $i \in \{m+1,\cdots,n\}$, {\color{black}
we adopt the OMA transmission strategy, and}
$s_i$ is designed as $s_i=s_{2i}$, i.e., $s_i$ only contains the
message for user $2$.
The corresponding information rate of $s_{2i}$, $i \in \{m+1,\cdots,n\}$,  can be expressed as
$
R_{2i}=\log \left(  1+\frac{P }{t^2 d_2^\tau N_0} \right).
$
When $i \in \{r+1,\cdots,m\}$, $s_i$ is observed by both  users.
The observations of $s_i$, $i \in \{r+1,\cdots,m\}$, at the two users
can be expressed as
\begin{eqnarray}
y_{1i}=\frac{\sqrt{P}}{t\sqrt{d_1^\tau} }
\alpha_{i-r} s_i +n_{1i}
\quad \text{and}  \quad
y_{2i}=\frac{\sqrt{P}}{t\sqrt{d_2^\tau} }
\beta_{i-r} s_i +n_{2i},
\end{eqnarray}
respectively. Again,
by applying F-NOMA, when
$\frac{\alpha_{i-r}}{\sqrt{d_1^\tau}} >\frac{\beta_{i-r}}{\sqrt{d_2^\tau}}$, $i \in \{r+1,\cdots,m\}$, i.e.,
$w_{i-r}^2 > \frac{d_1^\tau}{d_2^\tau}$,
$s_i$, $i \in \{r+1,\cdots,m\}$, is designed as $s_i=l_2 s_{1i}+l_1 s_{2i}$, where $s_{1i}$
and $s_{2i}$ are the information bearing messages for user $1$ and user $2$, respectively.
When $\frac{\alpha_{i-r}}{\sqrt{d_1^\tau}}  \leq \frac{\beta_{i-r}}{\sqrt{d_2^\tau}}$, $i \in \{r+1,\cdots,m\}$, i.e.,
$w_{i-r}^2  \leq \frac{d_1^\tau}{d_2^\tau}$,
$s_i$, $i \in \{r+1,\cdots,m\}$, is designed as $s_i=l_1 s_{1i}+l_2 s_{2i}$.
{\color{black}Note that $\mathcal{E} \{  |s_{i}|^2 \}=\mathcal{E} \{  |s_{1i}|^2 \}=\mathcal{E} \{  |s_{2i}|^2 \}=1$,
$i \in \{r+1,\cdots,m\}$.}
Therefore,
when $w_{i-r}^2 > \frac{d_1^\tau}{d_2^\tau}$, $i \in \{r+1,\cdots,m\}$, SIC is carried out at user $1$  and
the information rates of $s_{1i}$ and $s_{2i}$, $i \in \{r+1,\cdots,m\}$, can be expressed as
\begin{eqnarray}
\label{m<n1>2R}
R_{1i}=\log \left(  1+\frac{P \alpha_{i-r}^2  l_2^2}{ t^2 d_1^\tau N_0} \right)
\quad \text{and}  \quad
R_{2i}=\log \left(  1+\frac{P \beta_{i-r}^2  l_1^2}{P \beta_{i-r}^2  l_2^2+ t^2 d_2^\tau N_0} \right),
\end{eqnarray}
respectively.
When
$w_{i-r}^2 \leq \frac{d_1^\tau}{d_2^\tau}$,  SIC is carried by user $2$  and
the information rates of $s_{1i}$ and $s_{2i}$, $i \in \{r+1,\cdots,m\}$, can be expressed as
\begin{eqnarray}
\label{m<n1<2R}
R_{1i}=\log \left(  1+\frac{P \alpha_{i-r}^2  l_1^2}{P \alpha_{i-r}^2  l_2^2+ t^2 d_1^\tau N_0} \right)
\quad \text{and}  \quad
R_{2i}=\log \left(  1+\frac{P \beta_{i-r}^2  l_2^2}{ t^2 d_2^\tau N_0} \right),
\end{eqnarray}
respectively.
{\color{black}
Note that, similar to  \eqref{m>n1>2R} and  \eqref{m>n1<2R}, \eqref{m<n1>2R} and  \eqref{m<n1<2R}
provide
the achievable instantaneous
rates of the users  for given user channel matrices
$\mathbf{H}_1$ and $\mathbf{H}_2$.}

\subsubsection{The case when   $2m  \leq n$}
\label{2m>n}
{\color{black}
As shown in Section~\ref{GSVDdef},
when   $2m  \leq n$,
$\mathbf{\Sigma}_1$ and $\mathbf{\Sigma}_2$  can be expressed as in \eqref{mmn2}}.
So, when $2m  \leq n$, after applying detection matrices  $\mathbf{U}$ and  $\mathbf{V}$,
user $1$ and user $2$ observe
\begin{eqnarray}
\label{dd13}
\mathbf{U} \mathbf{y}_1=\frac{\sqrt{P}}{t\sqrt{d_1^\tau} }\left[ \begin{array}{c}
s_1\\
\vdots\\
s_m
\end{array} \right] +\mathbf{n}_1'
\quad \text{and}  \quad
\mathbf{V} \mathbf{y}_2=\frac{\sqrt{P}}{t\sqrt{d_2^\tau} }\left[ \begin{array}{c}
s_{n-m+1}\\
\vdots\\
s_n
\end{array} \right] +\mathbf{n}_2',
\end{eqnarray}
respectively.
{\color{black}
Again,  we assume that the
base station allocates the same power  to each
parallel SISO channel, i.e.,
$\mathcal{E} \{  |s_{i}|^2 \}=1$, $i \in \{1,\cdots,m,n-m+1,\cdots,n\}$}.
For $i \in \{1,\cdots,m\}$,
$s_i$  is  observed by user $1$ only, and
{\color{black}the observations of $s_i$, $i \in \{1,\cdots,m\}$, at the two users
can be expressed as
\begin{eqnarray}
y_{1i}=\frac{\sqrt{P}}{t\sqrt{d_1^\tau} }
s_i +n_{1i}
\quad \text{and}  \quad
y_{2i}=0,
\end{eqnarray}
respectively.
}So, when $i \in \{1,\cdots,m\}$,
{\color{black}
we adopt the OMA transmission strategy, and} $s_i$ is designed as $s_i=s_{1i}$, i.e., $s_i$ only contains the
information bearing message for user $1$. Therefore,
the corresponding information rate of $s_{1i}$, $i \in \{1,\cdots,m\}$,  can be expressed as $
R_{1i}=\log \left(  1+\frac{P }{t^2 d_1^\tau N_0} \right).
$ When $i \in \{n-m+1,\cdots,n\}$, $s_i$  is  observed by user $2$ only, and
{\color{black}the observations of $s_i$, $i \in \{n-m+1,\cdots,n\}$, at the two users
can be expressed as
\begin{eqnarray}
y_{1i}=0
\quad \text{and}  \quad
y_{2i}=\frac{\sqrt{P}}{t\sqrt{d_2^\tau} }
s_i +n_{2i},
\end{eqnarray}
respectively.}
Hence,  when $i  \in \{n-m+1,\cdots,n\}$, {\color{black}
we also adopt the OMA transmission strategy, and}
$s_i$ is designed as $s_i=s_{2i}$, i.e., $s_i$  only contains the
information bearing message for user $2$.
The corresponding information rate of $s_{2i}$, $i \in \{n-m+1,\cdots,n\}$,  can be expressed as
$
R_{2i}=\log \left(  1+\frac{P }{t^2 d_2^\tau N_0} \right).
$
For $i \in \{m+1,\cdots,n-m\}$,
$s_i$  is observed neither by
user $1$ nor by  user $2$.
So, $s_i$, $i \in \{m+1,\cdots,n-m\}$, is set as $s_i=0$ in order
to save  transmit power.

   \vspace{-1em}

\section{Performance Analysis}
\label{perfom}
In this section,  we  evaluate the performance of the proposed GSVD-NOMA protocol
for the  asymptotic case, where
the numbers of  transmit and receive
antennas approach infinity while the ratio of the numbers of  receive and transmit
antennas remains constant,
i.e.,  $\eta=\frac{m}{n}$ is
constant\footnote{{\color{black}
Please note that, in this section,
the performance analysis
of the GSVD-NOMA system encompasses
all possible cases of $m$ and $n$, i.e., $2m \leq n$, $m<n<2m$, and $m \geq n$.
The case $m \leq n$ applies
when the users have fewer antennas than the
base station, which is an expected scenario in the Internet of Things and  cellular  networks.
Furthermore, the
case $m \geq n$ is  applicable to e.g.
small cells
\cite{peng2015recent} and  cloud radio access networks (C-RANs) \cite{mobile2011c}, where low-cost base stations are deployed with high density
and these base stations are expected to have capabilities
similar to those of advanced smart phones
and tablets.
}}.
To this end,
we  first characterize {\color{black}the limiting distribution} of
the squared  generalized singular values $w_i^2=\alpha_i^2/\beta_i^2$ in the considered asymptotic regime.
Then, we study  the characteristics of power normalization factor $t$.
Finally, we calculate the normalized  average individual rates of the two users.
The derived analytical results can be applied when the base station and the users have large numbers of antennas. {\color{black}
For example,  in heterogenous networks,  a macro base station may communicate with two
micro base stations by using GSVD-NOMA. In this case, it
is reasonable to assume that both the transmitter and the receivers have large numbers of antennas.
Furthermore, the numerical results in Section~\ref{se5} reveal that
when the base station and the users have moderate numbers of antennas  (e.g. $m=2, n=5$),
the derived analytical results provide still
accurate approximations.}
Therefore, our asymptotic results provide
insight into the performance achieved by the proposed GSVD-NOMA scheme
for the realistic case where the base station and the users have  moderate numbers of antennas.

\vspace{-1em}

\subsection{{\color{black}The limiting distribution} of the squared generalized singular values}
As shown in Section~\ref{GSVDNOMA}, the rates of the users depend on
the  squared generalized singular values of the two users' channel matrices.
Therefore, in order to calculate {\color{black} the normalized  average individual rates} of the two users,
we first need to characterize {\color{black}the limiting distribution} of
{\color{black} the squared generalized singular values}.

Assuming that the elements of $\mathbf{H}_1$ and $\mathbf{H}_2$ are i.i.d. complex Gaussian random variables with
zero mean and unit variance, the distribution of {\color{black} the squared  generalized singular values} $w_i^2$ can be characterized as
follows.
{\color{black}
First, let us define the empirical distribution function (e.d.f.)
of $k$ random variables $v_i$, $i \in \{1,\cdots,k\}$, as
$\mathbf{F}_{v_i}^k(x)$, where
\begin{eqnarray}
\mathbf{F}_{v_i}^k(x)=\frac{1}{k}\sum_{j=1}^{k}1\{v_j\leq x\},
\end{eqnarray}
and $1\{\cdot\}$ is the indicator function.
Next,
we define  function $f_{y, y'}(x)$  as follows:
\begin{eqnarray}
\label{fy}
f_{y, y'}(x)=
\left\{ \begin{array}{ll}
\frac{ (1-y')\sqrt{ \left( x-    \left(    \frac{1- g(y,y') }{1-y'}   \right)^2      \right)
\left(       \left(    \frac{1+ g(y,y') }{1-y'}   \right)^2       -x                \right) }
}{2 \pi x(x y' +y)},
 &    \left(    \frac{1-  g(y,y') }{1-y'}   \right)^2<x< \left(    \frac{1+ g(y,y') }{1-y'}   \right)^2 \\
0, & \textrm{otherwise}
\end{array} \right.
\end{eqnarray}
where $x$ is the argument of the  function, $y$ and $y'$ are two parameters of the function, and
$g(y,y')=\sqrt{1-(1-y)(1-y')}$.

Then, {\color{black} equipped with there definitions}, we can characterize the distribution of  the squared  generalized singular values $w_i^2$ as in the following theorem.
\begin{theorem}
\label{GSVDvalue}
Suppose that $\mathbf{H}_1$ and $\mathbf{H}_2$ are two  $m \times n$
matrices whose elements are i.i.d. complex Gaussian random variables
with
zero mean and unit variance
and their GSVD is defined as in \eqref{GSVDFORM}.
\begin{itemize}
\item{
When $m\geq n$, almost surely, $\mathbf{F}_{w_i}^n(x)$, the e.d.f. of their squared generalized singular values $w_i^2$, $i \in \{1,\cdots,n\}$,
converges, as $m,n \to \infty$ with $\frac{m}{n}\to \eta$, to
a nonrandom cumulative distribution function (c.d.f.) $\mathbf{F}_{w_i}(x)$, whose
probability density function (p.d.f.) is $f_{\frac{1}{\eta}, \frac{1}{\eta}}(x)$.}
\item{
When $m<n<2m$, almost surely,  $\mathbf{F}_{w_i}^{2m-n}(x)$, the e.d.f. of their  squared generalized singular values $w_i^2$, $i \in \{1,\cdots, 2m-n\}$,
converges, as $m,n \to \infty$ with $\frac{m}{n}\to \eta$, to
a nonrandom c.d.f. $\mathbf{F}_{w_i}(x)$, whose
p.d.f. is  $\frac{\eta}{(2\eta-1)^2}f_{\frac{\eta}{2\eta-1}, \eta}\left(\frac{x}{2\eta-1}\right)$.
}
\item{
When $2m \leq n$,  $\mathbf{\Sigma}_1$ and $\mathbf{\Sigma}_2$  are
deterministic and
given by  $[\mathbf{I}_m \quad \mathbf{O}_{m \times (n-m)}]$ and
$[\mathbf{O}_{m \times (n-m)} \quad \mathbf{I}_m]$, respectively.
}
\end{itemize}
\end{theorem}
\begin{IEEEproof}
See Appendix A.
\end{IEEEproof}
}


   \vspace{-1em}
\subsection{The value of the power normalization factor $t$}
As shown in Section~\ref{GSVDNOMA}, the power normalization factor $t$
is
a key parameter affecting
the rates of the users.
In this subsection, we characterize  the value of this power normalization factor.
{\color{black}
Recall that $t^2$
can be expressed as
$t^2= \mathcal{E} \{ \text{trace} (\mathbf{Q}
\mathbf{s} \mathbf{s}^H \mathbf{Q}^H) \}$, where  $\mathbf{Q}$ is the GSVD decomposition matrix defined as in \eqref{GSVDFORM}, and $\mathbf{s}$ is the information bearing symbol vector.
{\color{black}
Moreover, as shown in Section~\ref{GSVDNOMA},
$\mathcal{E} \{\mathbf{s} \mathbf{s}^H \}$
can be expressed  as
$\mathcal{E} \{\mathbf{s} \mathbf{s}^H \}=
\left\{ \begin{array}{ll}
\mathbf{I}_n  &    2m \geq n \\
\mathbf{B} & 2m < n
\end{array} \right.,$
where  $n \times n$ matrix $\mathbf{B}$
= $\text{diag} \left[ \mathbf{I}_m, \mathbf{O}_{(n-2m) \times (n-2m)}, \mathbf{I}_m  \right]$.
Thus,  $t^2$ can be further expressed as follows:
\begin{eqnarray}
t^2 = \mathcal{E} \{ \text{trace} (\mathbf{Q}
\mathbf{s} \mathbf{s}^H \mathbf{Q}^H) \}=
\left\{ \begin{array}{ll}
\mathcal{E} \{\text{trace} (\mathbf{Q}\mathbf{Q}^H)\} &    2m \geq n \\
 \mathcal{E} \{\text{trace} (\mathbf{Q}
\mathbf{B} \mathbf{Q}^H)\} & 2m < n
\end{array} \right..
\end{eqnarray}}
{\color{black} Moreover, let us define $\mathbf{H}=\left(\begin{array}{c}
\mathbf{H}_1
\\
\mathbf{H}_2
\end{array}\right)$ and $\eta=\frac{m}{n}$.}
Then,  the value of $t^2$ is discussed as follows.

\begin{itemize}
\item{For the case of $2m > n$, the value of $t^2$ is given by the following theorem.
\begin{theorem}
\label{Q}
Assume that  $\mathbf{H}_1$ and $\mathbf{H}_2$ are two  $m \times n$
matrices whose elements are i.i.d. complex Gaussian random variables with
zero mean and unit variance and their GSVD is defined as  in   \eqref{GSVDFORM}.
Then, for $2m > n$, $t^2=\mathcal{E} \{\text{trace} (\mathbf{Q}\mathbf{Q}^H)\}
=\mathcal{E} \{\text{trace}((\mathbf{H}^H\mathbf{H})^{-1})\}=\frac{1}{2\eta -1}$.
\end{theorem}
\begin{IEEEproof}
See Appendix B.
\end{IEEEproof}}

\item{For the case of $2m < n$, the value of $t^2$ is given by the following theorem.
\begin{theorem}
\label{Q33}
Assume that  $\mathbf{H}_1$ and $\mathbf{H}_2$ are two  $m \times n$
matrices whose elements are i.i.d. complex Gaussian random variables with
zero mean and unit variance and their GSVD is defined as   in  \eqref{GSVDFORM}.
Then, for $2m < n$, $t^2=\mathcal{E} \{\text{trace} (\mathbf{Q}\mathbf{B}\mathbf{Q}^H)\}
=\mathcal{E} \{\text{trace}((\mathbf{H}\mathbf{H}^H)^{-1})\}=\frac{1}{1/(2\eta) -1}$.
\end{theorem}
\begin{IEEEproof}
See Appendix C.
\end{IEEEproof}}

{\color{black}
\item{For the case of $2m = n$,  from Appendix B, it can be shown that  $t^2= \mathcal{E} \{\text{trace} (\mathbf{Q}\mathbf{Q}^H)\}=\mathcal{E} \{\text{trace} ((\mathbf{H}^H\mathbf{H})^{-1}) \}$.
Note that the elements of $\mathbf{H}$ are i.i.d. complex Gaussian random variables
with
zero mean and unit variance. From \cite{tulino2004foundations}, it is easy to show that
when $2m=n$, $\mathcal{E} \{\text{trace} ((\mathbf{H}^H\mathbf{H})^{-1}) \}$ approaches infinity.
Therefore, when $n=2m$, the average power of the GSVD precoding matrix, $t^2$, approaches  infinity.
Thus, when $n=2m$, the long-term power constraint is not applicable, and in practice  more sophisticated precoding schemes
based on an instantaneous power constraint
should be applied at the base
station. {\color{black} Studying such precoders is beyond the scope of this paper.}
}}
\end{itemize}}


\subsection{ The normalized  average  individual rates of the two users}
\label{average rate}
In this subsection, we focus on
 the normalized  average  individual rates of the two users.

\subsubsection{The case of  $m \geq n$}
As shown in Section~\ref{m>n}, when
$m \geq n$,  $s_{1i}$
and $s_{2i}$, $i \in \{1,\cdots, n\}$,  are broadcasted by the base station,
where $s_{1i}$
and $s_{2i}$ are the information bearing messages for user $1$ and user $2$, respectively.
The instantaneous information rates of $s_{1i}$ and $s_{2i}$ can be expressed as in \eqref{m>n1>2R}
and \eqref{m>n1<2R}.
Recall that  {\color{black} the  squared generalized singular values} are defined as
$w_i^2=\alpha_i^2/\beta_i^2$, $i \in \{1,\cdots, n\}$.
From the fact that
$\alpha_i^2 + \beta_i^2 = 1$, we can obtain that
$\alpha_i^2=\frac{w_i^2}{1+w_i^2}$ and $\beta_i^2=\frac{1}{1+w_i^2}$.
Therefore, when
$m \geq n$,  substituting $\alpha_i^2=\frac{w_i^2}{1+w_i^2}$ and $\beta_i^2=\frac{1}{1+w_i^2}$
into  \eqref{m>n1>2R},
we can show that
when
$w_i^2 > \frac{d_1^\tau}{d_2^\tau}$,
the instantaneous information rates of $s_{1i}$ and $s_{2i}$ can be expressed as follows:
\begin{eqnarray}
\label{r1}
&&R_{1i}=\log \left(  1+\frac{P \alpha_i^2  l_2^2}{ t^2 d_1^\tau N_0} \right)  =  \log \left(  1+\frac{P w_i^2  l_2^2}{ t^2 d_1^\tau N_0 (1+w_i^2)} \right)
\end{eqnarray}
and
\begin{eqnarray}
\label{r2}
&&R_{2i}=\log \left(  1+\frac{P \beta_i^2  l_1^2}{P \beta_i^2  l_2^2+ t^2 d_2^\tau N_0} \right)   =\log \left(  1+\frac{P   l_1^2}{P
 l_2^2+ t^2 d_2^\tau N_0 (1+w_i^2)} \right).
\end{eqnarray}
Furthermore,  substituting $\alpha_i^2=\frac{w_i^2}{1+w_i^2}$ and $\beta_i^2=\frac{1}{1+w_i^2}$
into  \eqref{m>n1<2R},
we can show that when
$w_i^2 \leq \frac{d_1^\tau}{d_2^\tau}$,
the instantaneous information rates of $s_{1i}$ and $s_{2i}$ can be expressed as as follows:
\begin{eqnarray}
\label{r3}
&&R_{1i}=\log \left(  1+\frac{P \alpha_i^2  l_1^2}{P \alpha_i^2  l_2^2+ t^2 d_1^\tau N_0} \right)   = \log \left(  1+\frac{P w_i^2  l_1^2}{P w_i^2  l_2^2+ t^2 d_1^\tau N_0 (1+w_i^2)} \right)
\end{eqnarray}
and
\begin{eqnarray}
\label{r4}
&&R_{2i}=\log \left(  1+\frac{P \beta_i^2  l_2^2}{ t^2 d_2^\tau N_0} \right)  = \log \left(  1+\frac{P  l_2^2}{ t^2 d_2^\tau N_0 (1+w_i^2)} \right).
\end{eqnarray}
So when $m \geq n$, $m,n \to \infty$ with $\frac{m}{n}\to \eta$,
{\color{black} the normalized
average individual rates} of the two users
can be characterized as in the following corollary.
\begin{corollary}
\label{cor1}
When $m \geq n$, $m,n \to \infty$ with $\frac{m}{n}\to \eta$, {\color{black} the normalized average individual rates} of user $1$ and user $2$  can be expressed as follows:
\begin{eqnarray}
R_1&=&\frac{1}{\eta}  \left(    C\left(\frac{d_1^\tau}{d_2^\tau},B, \frac{d_1^\tau N_0}{d_1^\tau N_0 + P l_2^2 (2 \eta -1)}\right)
-  C\left(\frac{d_1^\tau}{d_2^\tau},B, 1\right) \right.
 \\   &&  \left.
+\log \left(  \frac{  d_1^\tau N_0 +P l_2^2 (2 \eta -1)}{  d_1^\tau N_0 } \right) D\left(\frac{d_1^\tau}{d_2^\tau},B \right)
+ C\left(A,\frac{d_1^\tau}{d_2^\tau}, \frac{d_1^\tau N_0}{d_1^\tau N_0 + P  (2 \eta -1)}\right)
\right.
\nonumber \\   &&  \nonumber  \left.
- C\left(A,\frac{d_1^\tau}{d_2^\tau}, \frac{d_1^\tau N_0}{d_1^\tau N_0 + P l_2^2  (2 \eta -1)}\right)
+\log \left(  \frac{  d_1^\tau N_0 +P (2 \eta -1)}{  d_1^\tau N_0 +P l_2^2 (2 \eta -1)} \right) D\left(A,\frac{d_1^\tau}{d_2^\tau} \right)
\right)
\end{eqnarray}
and
\begin{eqnarray}
R_{2}&=& \frac{1}{\eta}  \left(
C\left(\frac{d_1^\tau}{d_2^\tau},B, 1+\frac{P  (2 \eta -1)}{d_2^\tau N_0 }\right)
-
C\left(\frac{d_1^\tau}{d_2^\tau},B, 1+\frac{P l_2^2  (2 \eta -1)}{d_2^\tau N_0 }\right)
\right.
\nonumber \\   &&
\left.
+C\left(A, \frac{d_1^\tau}{d_2^\tau}, \frac{P l_2^2  (2 \eta -1)+d_2^\tau N_0}{d_2^\tau N_0 }\right)
-C\left(A,\frac{d_1^\tau}{d_2^\tau}, 1\right)
\right),
\end{eqnarray} where $A=\left(    \frac{1-\sqrt{1-(1-1/\eta)(1-1/\eta)} }{1-1/\eta}   \right)^2$
, $B=\left(    \frac{1+\sqrt{1-(1-1/\eta)(1-1/\eta)} }{1-1/\eta}   \right)^2$
,
 $C(y_1,y_2,y_3)= \int_{y_1}^{y_2} \log(x+y_3) f_{\frac{1}{\eta}, \frac{1}{\eta}}(x) dx$,
and $D(y_1,y_2)= \int_{y_1}^{y_2}  f_{\frac{1}{\eta}, \frac{1}{\eta}}(x) dx$.
\end{corollary}
\begin{IEEEproof}{\color{black}
See Appendix D.}
\end{IEEEproof}

\subsubsection{The case when   $m<n<2m$}
As shown in Section~\ref{m<n2m>n},
when
$m<n<2m$, $s_{1i}$, $i \in \{1,\cdots, m\}$,
and $s_{2i}$, $i \in \{r+1,\cdots, n\}$,  are broadcasted by the base station,
where $r=n-m$, and $s_{1i}$
and $s_{2i}$ are the information bearing  messages for user $1$ and user $2$, respectively.
The instantaneous information rate of $s_{1i}$, $i \in \{1,\cdots, r\}$,  is given by
$
R_{1i}=\log \left(  1+\frac{P }{t^2 d_1^\tau N_0} \right).
$
The instantaneous information rate of $s_{2i}$, $i \in \{m+1,\cdots, n\}$,  is given by
$
R_{2i}=\log \left(  1+\frac{P }{t^2 d_2^\tau N_0} \right).
$
When $i \in \{r+1,\cdots, m\}$, $s_{1i}$
and $s_{2i}$ are observed by both users.
The instantaneous information rates of $s_{1i}$
and $s_{2i}$, $i \in \{r+1,\cdots, m\}$, are given by  \eqref{m<n1>2R}
and \eqref{m<n1<2R}.
Recall that when $m<n<2m$, {\color{black} the squared  generalized singular values} are defined as
$w_i^2=\alpha_i^2/\beta_i^2$, $i \in \{1,\cdots, 2m-n\}$. As
$\alpha_i^2 + \beta_i^2 = 1$, {\color{black} we obtain}
$\alpha_i^2=\frac{w_i^2}{1+w_i^2}$ and $\beta_i^2=\frac{1}{1+w_i^2}$.
Therefore, when
$m<n<2m$,  from \eqref{m<n1>2R} and \eqref{m<n1<2R},
{\color{black}it can be shown that}
when
$w_{i-r}^2 > \frac{d_1^\tau}{d_2^\tau}$, $i \in \{r+1,\cdots, m\}$,
the instantaneous information rates of $s_{1i}$ and $s_{2i}$ can be expressed as follows:
\begin{eqnarray}
\label{r11}
R_{1i}&=&\log \left(  1+\frac{P \alpha_{i-r}^2  l_2^2}{ t^2 d_1^\tau N_0} \right)=\log \left(  1+\frac{P w_{i-r}^2  l_2^2}{ t^2 d_1^\tau N_0 (1+ w_{i-r}^2)} \right)
\end{eqnarray}
and
\begin{eqnarray}
\label{r21}
R_{2i}&=&\log \left(  1+\frac{P \beta_{i-r}^2  l_1^2}{P \beta_{i-r}^2  l_2^2+ t^2 d_2^\tau N_0} \right)=
\log \left(  1+\frac{P   l_1^2}{P
 l_2^2+ t^2 d_2^\tau N_0 (1+w_{i-r}^2)} \right).
\end{eqnarray}
On the other hand,  when
$w_{i-r}^2 \leq \frac{d_1^\tau}{d_2^\tau}$, $i \in \{r+1,\cdots, m\}$,
the instantaneous information rates of $s_{1i}$ and $s_{2i}$ can be expressed as  follows:
\begin{eqnarray}
\label{r31}
R_{1i}&=&\log \left(  1+\frac{P \alpha_{i-r}^2  l_1^2}{P \alpha_{i-r}^2  l_2^2+ t^2 d_1^\tau N_0} \right)
 = \log \left(  1+\frac{P w_{i-r}^2  l_1^2}{P w_{i-r}^2  l_2^2+ t^2 d_1^\tau N_0 (1+w_{i-r}^2)} \right)
\end{eqnarray}
and
\begin{eqnarray}
\label{r41}
R_{2i}=\log \left(  1+\frac{P \beta_{i-r}^2  l_2^2}{ t^2 d_2^\tau N_0} \right)
= \log \left(  1+\frac{P  l_2^2}{ t^2 d_2^\tau N_0 (1+w_{i-r}^2)} \right).
\end{eqnarray}
Thus, when $m<n<2m$, $m,n \to \infty$ with $\frac{m}{n}\to \eta$,
the normalized
average individual rates of the two users
can be characterized as in the following corollary.
\begin{corollary}
When $m<n<2m$, $m,n \to \infty$ with $\frac{m}{n}\to \eta$, {\color{black} the normalized
average  individual rates} of  user $1$ and user $2$  can be expressed as follows:
\begin{eqnarray}
R_1&=&\left(2-\frac{1}{\eta}\right)  \left(    G\left(\frac{d_1^\tau}{d_2^\tau}, F, \frac{d_1^\tau N_0}{d_1^\tau N_0 + P l_2^2 (2 \eta -1)}\right)
-  G\left(\frac{d_1^\tau}{d_2^\tau}, F, 1\right) \right.
 \\   &&  \left.
+\log \left(  \frac{  d_1^\tau N_0 +P l_2^2 (2 \eta -1)}{  d_1^\tau N_0 } \right) H\left(\frac{d_1^\tau}{d_2^\tau}, F \right)
+ G\left(E,\frac{d_1^\tau}{d_2^\tau}, \frac{d_1^\tau N_0}{d_1^\tau N_0 + P  (2 \eta -1)}\right)
\right.
\nonumber \\   &&  \nonumber  \left.
- G\left(E,\frac{d_1^\tau}{d_2^\tau}, \frac{d_1^\tau N_0}{d_1^\tau N_0 + P l_2^2  (2 \eta -1)}\right)
+\log \left(  \frac{  d_1^\tau N_0 +P (2 \eta -1)}{  d_1^\tau N_0 +P l_2^2 (2 \eta -1)} \right) H\left(E,\frac{d_1^\tau}{d_2^\tau} \right)
\right)
 \\   &&  \nonumber
 +\left(\frac{1}{\eta}-1\right)  \log \left(  1+\frac{P (2 \eta -1)}{ d_1^\tau N_0} \right)
\end{eqnarray}
and
\begin{eqnarray}
R_{2}&=& \left(2-\frac{1}{\eta}\right)  \left(
G\left(\frac{d_1^\tau}{d_2^\tau}, F, 1+\frac{P  (2 \eta -1)}{d_2^\tau N_0 }\right)
-
G\left(\frac{d_1^\tau}{d_2^\tau}, F, 1+\frac{P l_2^2  (2 \eta -1)}{d_2^\tau N_0 }\right)
\right.
\nonumber \\   &&
\left.
+G\left(E, \frac{d_1^\tau}{d_2^\tau}, \frac{P l_2^2  (2 \eta -1)+d_2^\tau N_0}{d_2^\tau N_0 }\right)
-G\left(E,\frac{d_1^\tau}{d_2^\tau}, 1\right)
\right)
\nonumber  \\   &&
 +\left(\frac{1}{\eta}-1\right)  \log \left(  1+\frac{P (2 \eta -1)}{ d_2^\tau N_0} \right),
\end{eqnarray}
where $E=(2\eta-1)\left(    \frac{1-\sqrt{1-(1-\eta/(2\eta-1))(1-\eta)} }{1-\eta}   \right)^2$, $F=(2\eta-1)\left(    \frac{1+\sqrt{1-(1-\eta/(2\eta-1))(1-\eta)} }{1-\eta}   \right)^2$, $G(y_1,y_2,y_3)= \int_{y_1}^{y_2} \log(x+y_3) \frac{\eta}{(2\eta-1)^2}f_{\frac{\eta}{2\eta-1}, \eta}(x/(2\eta-1)) dx$,
and $H(y_1,y_2)= \int_{y_1}^{y_2}  \frac{\eta}{(2\eta-1)^2}f_{\frac{\eta}{2\eta-1}, \eta}(x/(2\eta-1)) dx$.
\end{corollary}
\begin{IEEEproof}
{\color{black}
Following  steps similar to those in the  proof of Corollary \ref{cor1},
the normalized
average  individual rates of  the two users  can be obtained.}
\end{IEEEproof}

\subsubsection{The case when   $2m  <n$}
As shown in Section~\ref{2m>n},
when $2m  < n$,
$s_{1i}$
and $s_{2i}$, $i \in \{1,\cdots, n\}$,  are broadcasted by the base station.
The information rates of $s_{1i}$ and $s_{2i}$ can be expressed
as $
R_{1i}=\log \left(  1+\frac{P }{t^2 d_1^\tau N_0} \right)
$ and $
R_{2i}=\log \left(  1+\frac{P }{t^2 d_2^\tau N_0} \right)
$, respectively.
Moreover,
when  $2m < n$,
$t^2= \text{trace} (\mathbf{Q}\mathbf{s} \mathbf{s}^H\mathbf{Q}^H)=\text{trace} (\mathbf{Q} \mathbf{B} \mathbf{Q}^H)$.
{\color{black}
Theorem~\ref{Q33} obtains that
when $2m<n$, $t^2=\text{trace} (\mathbf{Q} \mathbf{B} \mathbf{Q}^H)$
converges, as $m,n \to \infty$ with $\frac{m}{n}\to \eta$, to $\frac{1}{1/(2\eta) -1}$.}
Thus, when $2m < n$,  $m,n \to \infty$ with $\frac{m}{n}\to \eta$, {\color{black} the normalized
average individual rates} of the two users   can be expressed as follows:
\begin{eqnarray}
R_1= \log \left(  1+\frac{P (1/(2\eta) -1)}{ d_1^\tau N_0} \right)
\quad \text{and}  \quad
R_{2}=\log \left(  1+\frac{P (1/(2\eta) -1)}{ d_2^\tau N_0} \right).
\end{eqnarray}

\vspace{-1em}

\section{Numerical Results} \label{se5}

{\color{black}
In this section, we first provide computer simulation results by focusing on the MIMO
downlink communication scenario with one base stations and two users,  where the base station and the users have large but finite numbers of antennas,
  to demonstrate the performance of the
proposed GSVD-NOMA scheme, and to verify the correctness of the developed analytical
results.
Then, we propose a hybrid NOMA scheme
to
extend
the proposed GSVD-NOMA scheme
to the MIMO
downlink communication scenario with
more than two users,
and provide numerical results
to  demonstrate
the performance of this hybrid NOMA scheme.
}

\begin{figure}
\begin{minipage}[tbp]{0.5\linewidth}
\centering
	\includegraphics[width=2.2in]{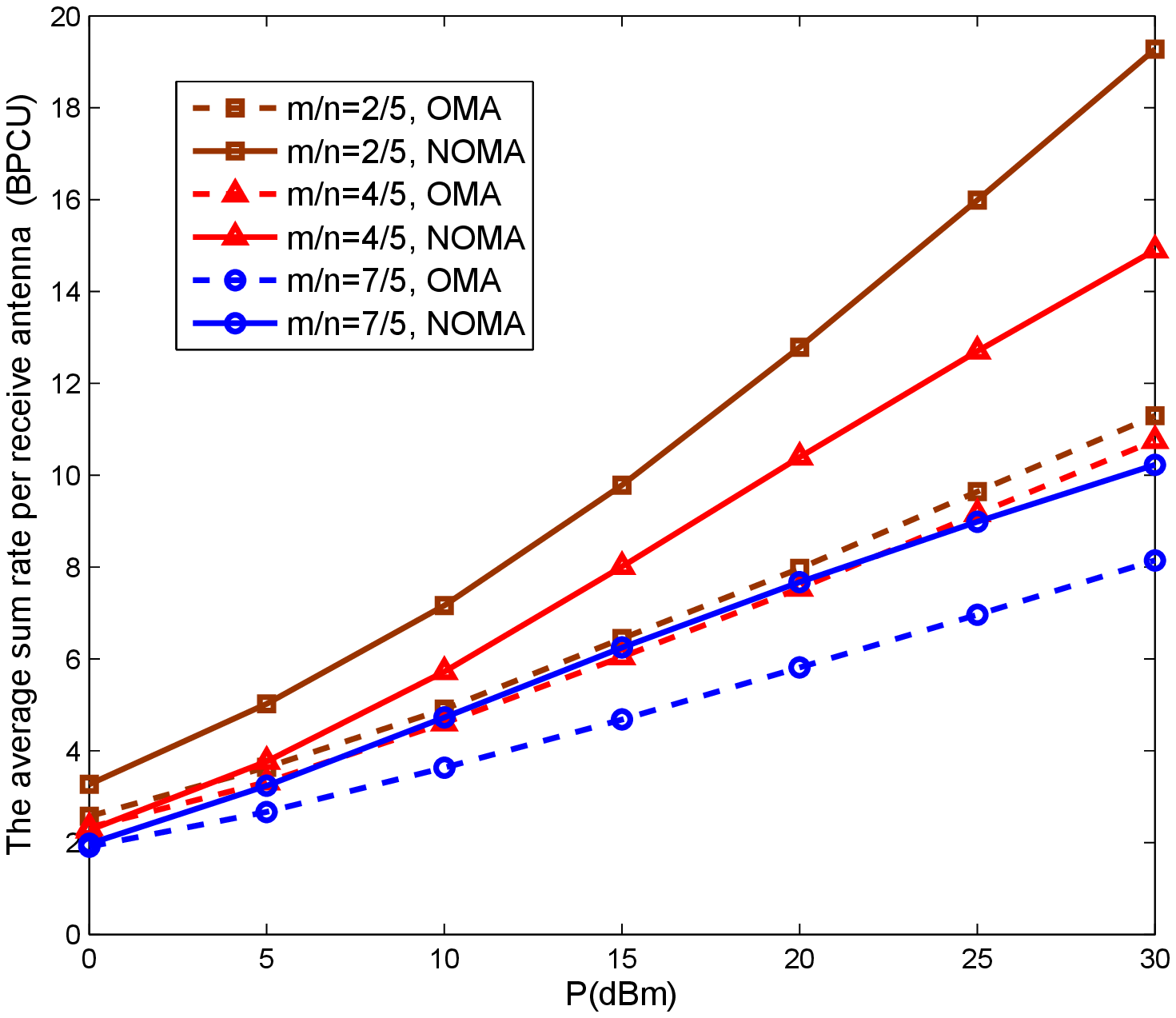}
\vspace{-2em}
    \caption{Comparing the normalized average sum rates achieved by   OMA and GSVD-NOMA. }
     \vspace{-1em}
    \label{Ratecom}
\end{minipage}
\begin{minipage}[tbp]{0.5\linewidth}
\centering
	\includegraphics[width=2.2in]{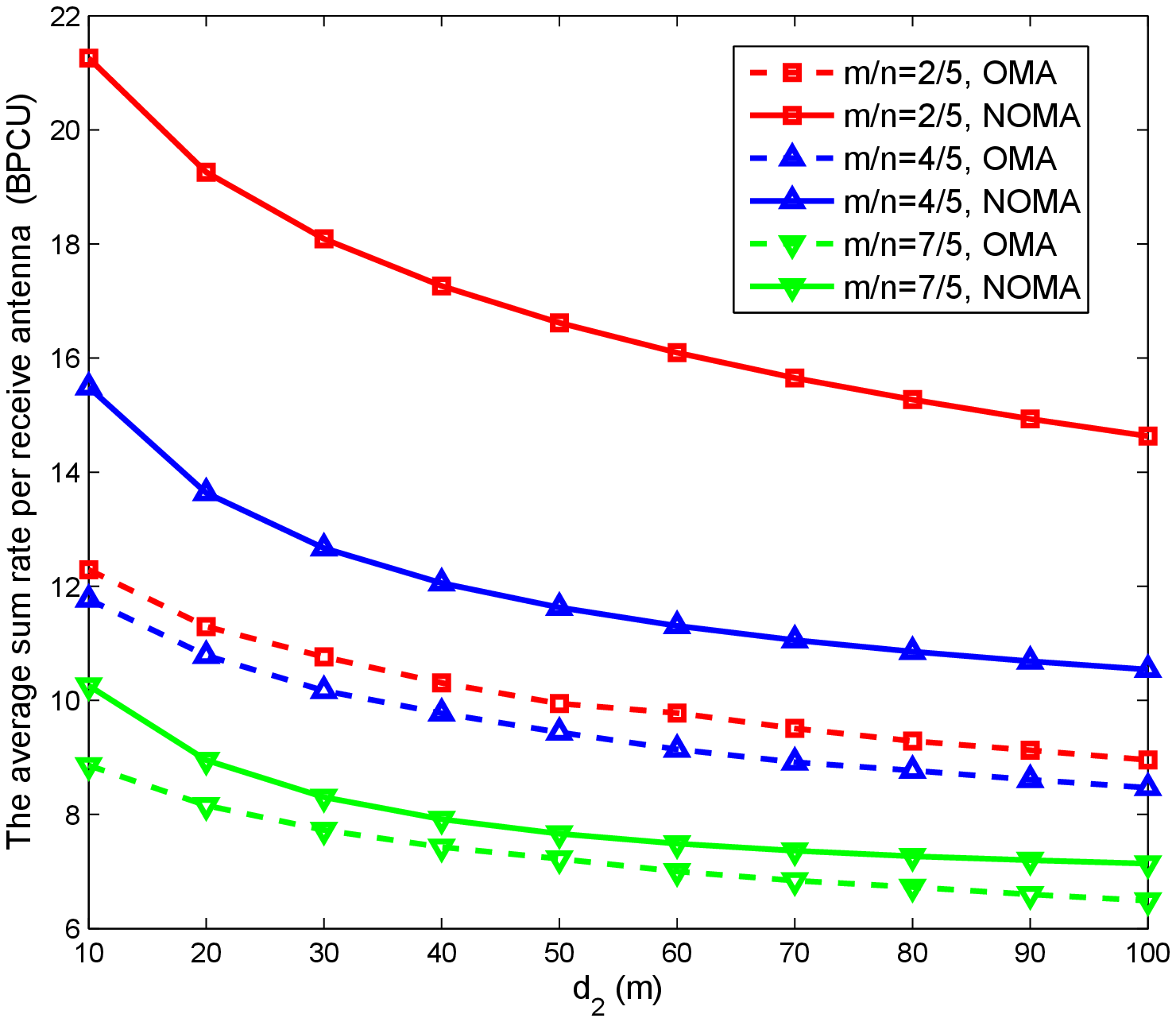}
\vspace{-2em}
 \caption{The normalized average sum rates achieved by   OMA and GSVD-NOMA versus distance $d_2$.}
     \vspace{-1em}
    \label{Ratecom2}
\end{minipage}
\end{figure}

In Fig. \ref{Ratecom},  we compare {\color{black} the normalized  average sum rates} achieved by OMA
and GSVD-NOMA
when $m=28$, $d_1=10$m, $d_2=100$m, $\tau=2$, $N_0=-35$dBm, and $l_2^2=0.2$.
Here, ``BPCU'' denotes bit per channel use.
From this figure, we observe that the proposed GSVD-NOMA scheme outperforms
conventional OMA\footnote{ {\color{black}
Here, the time division multiple access (TDMA) protocol is used as the
benchmark.
}} by a considerable margin. As
the transmit power $P$  increases, the performance gap between the two
protocols also increases, which confirms that the proposed GSVD-NOMA scheme
can exploit the spatial degrees of freedom of the channel more efficiently than OMA in the high SNR regime. Moreover,
when $\eta=\frac{m}{n}$ changes from $\frac{4}{5}$ to $\frac{2}{5}$, i.e., the number of the transmit antennas
increases,  {\color{black} the normalized  average sum rate} achieved by  OMA  stays  practically constant while
{\color{black} the normalized  average sum rate} achieved by  the proposed GSVD-NOMA scheme  increases considerably.
{\color{black}
This can be explained as follows,
with OMA,
the base station can perform SVD to
convert the $n \times m$ MIMO channel of each user into $k_1$ parallel SISO
channels}, {\color{black}where $k_1=\min\{m,n\}$.
Thus, in OMA systems,
when $n$ exceeds $m$,
the number of the parallel SISO
channels, $k_1$,
will stop growing,
which results in the
saturation of the normalized average
sum rate.
On the other hand,
as shown in Section~\ref{GSVDNOMA},
for the proposed GSVD-NOMA scheme,
the base station
decomposes the $n \times m$  MIMO channels of both users
into
$k_2$
parallel SISO channels,
where $k_2=\min\{2m,n\}$.} {\color{black}
Therefore,
when $n$ increases from $m$ to $2m$,
while $k_1$ will stay constant,
$k_2$  will  grow.
Hence, while
the normalized average
sum rate achieved by OMA   saturates quickly, that of  the proposed GSVD-NOMA scheme does not.}

Fig. \ref{Ratecom2} compares {\color{black} the normalized  average sum rates} achieved by  OMA and GSVD-NOMA
as a function of $d_2$
when $m=28$, $d_1=50$m, $\tau=2$, $P=30$dBm, $N_0=-35$dBm, and $l_2^2=0.2$.
From this figure, we observe that the proposed GSVD-NOMA   outperforms
conventional OMA for all considered values of $d_2$.
In fact, regardless of whether $d_1<d_2$ or $d_1>d_2$, the proposed GSVD-NOMA
achieves a higher sum rate than  conventional OMA.
Moreover, when the number of the transmit antennas increases,
the performance gap between the two
protocols increases from 1.5 BPCU  to 9.5 BPCU
for $d_2=10$m.

\begin{figure}
\begin{minipage}[tbp]{0.5\linewidth}
\centering
	\includegraphics[width=2.2in]{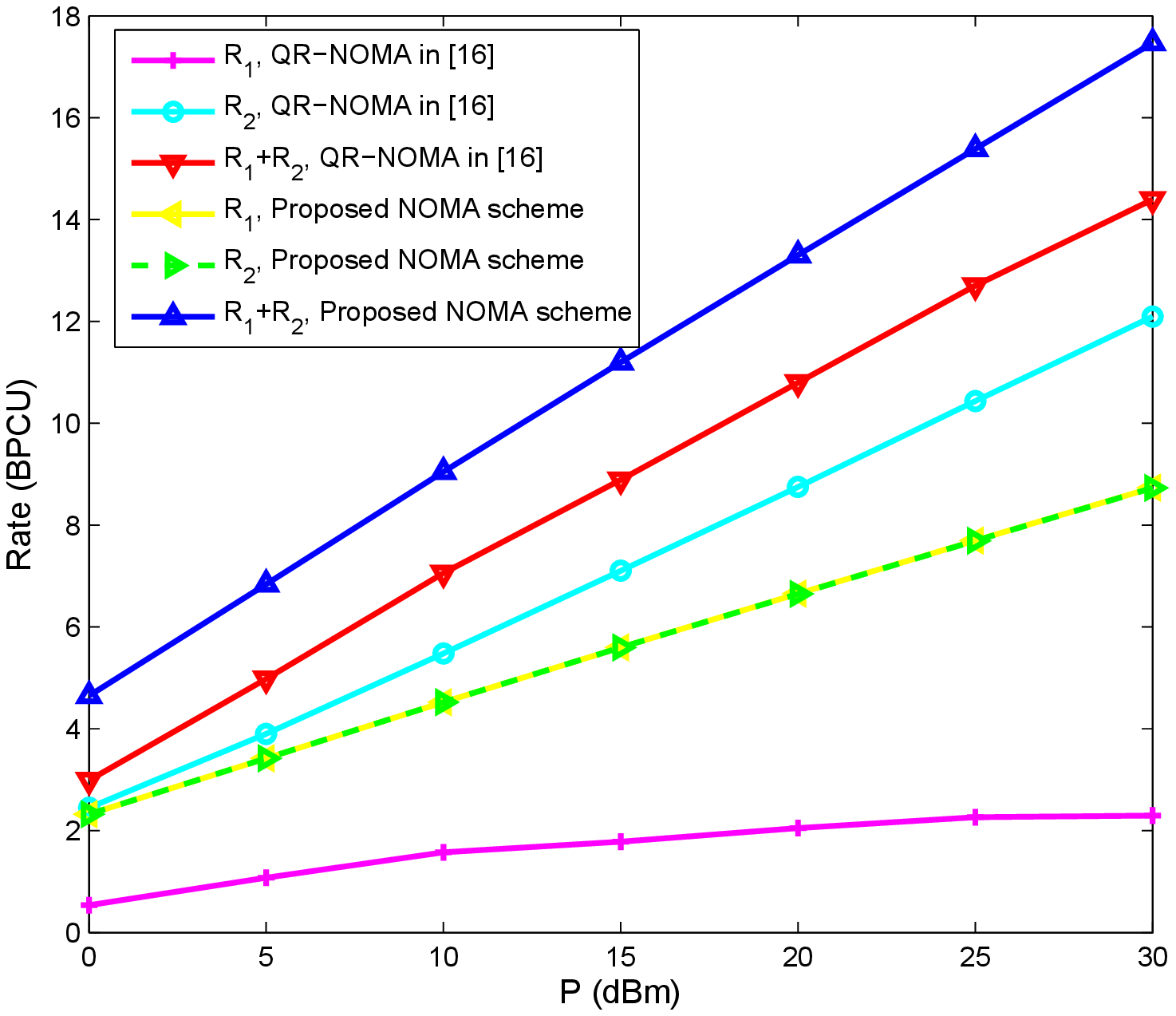}
\vspace{-2em}
    \caption{The normalized average rates achieved by the QR-NOMA scheme in \cite{ding2016mimo} and the proposed GSVD-NOMA scheme.}
     \vspace{-1em}
    \label{new1}
\end{minipage}
\begin{minipage}[tbp]{0.5\linewidth}
\centering
	\includegraphics[width=2.2in]{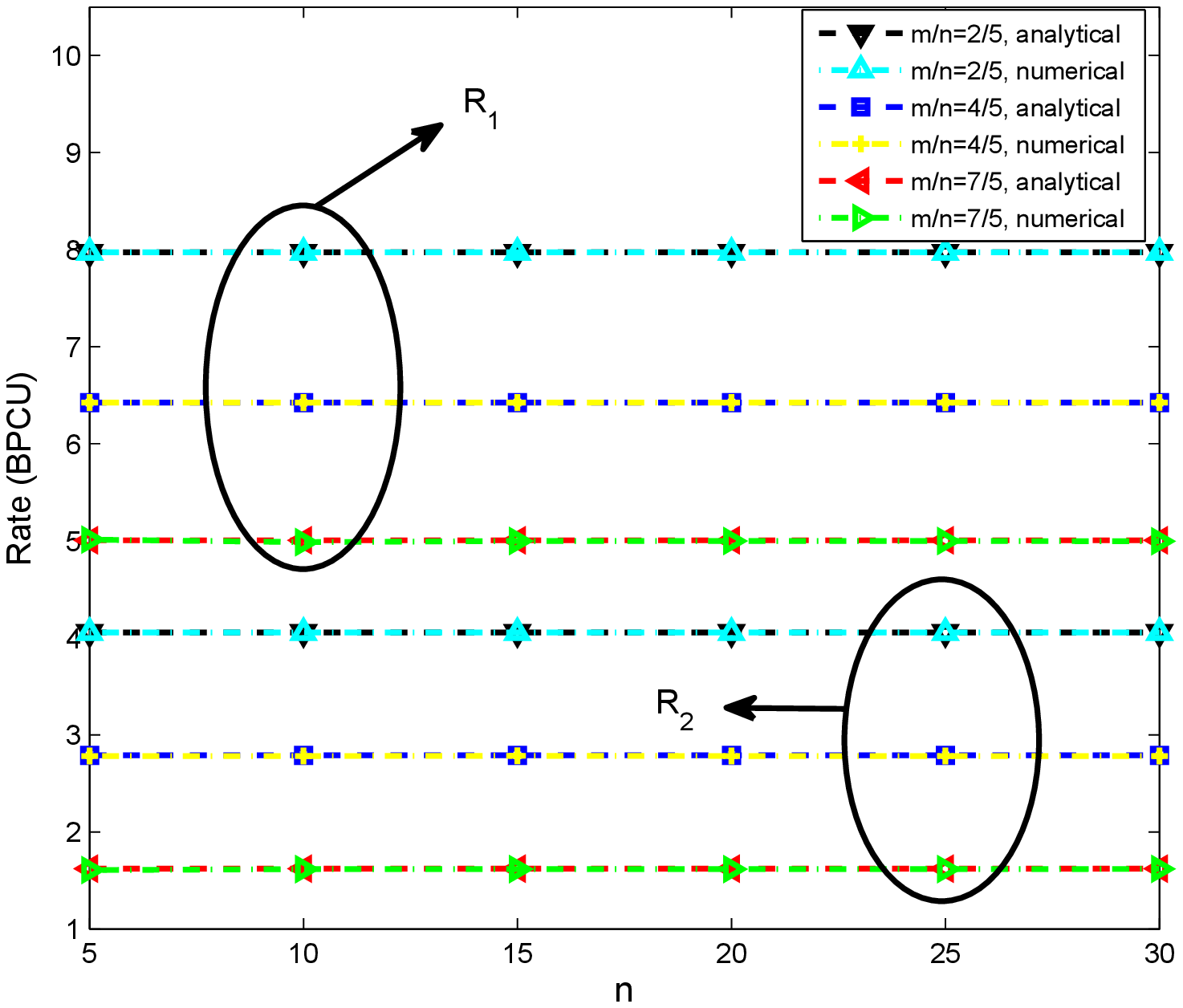}
     \vspace{-2em}
 \caption{{\color{black}The normalized average individual rates of the two users  versus $n$.}}
     \vspace{-1em}
    \label{R1R2}
\end{minipage}
\end{figure}

In Fig. \ref{new1},   the normalized  average rates achieved by
the proposed GSVD-NOMA scheme are compared to those achieved by
the QR-NOMA scheme in \cite{ding2016mimo}.
In this figure, it is assumed that $m=40$, $n=50$, $d_1=10$m, $d_2=10$m, $\tau=2$, $N_0=-35$dBm,
and $l_2^2=0.2$.
{\color{black}For  QR-NOMA  \cite{ding2016mimo},
the channel of one user will become
extremely weak,
which negatively affects the average sum rate.}
{\color{black}
Therefore,
as shown in Fig. \ref{new1},
when increasing the transmit power,
the average sum rate of the proposed GSVD-NOMA
scheme increases more rapidly than that of the
QR-NOMA scheme.}
{\color{black}
Moreover, for the proposed GSVD-NOMA scheme,
the normalized average individual rates
of the two users  are  identical, i.e., $R_1 = R_2$.
In contrast, for the
QR-NOMA scheme\cite{ding2016mimo}, $R_1$ is
 much greater than $R_2$.
Therefore,
the proposed GSVD-NOMA scheme ensures better fairness between the
two users than  QR-NOMA.}

{\color{black}
Fig. \ref{R1R2} shows the normalized   average individual rates of the two users
with $n$ increasing from $5$ to $30$,
when $d_1=10$m, $d_2=40$m, $\tau=2$,  $P=15$dBm, $N_0=-35$dBm, and $l_2^2=0.2$.
As can be seen from this
figure, when $m$ and $n$ is small (e.g. $m=2,n=5$), the
numerical results still coincide with  the analytical results perfectly,
which validates the analytical results  developed in Section~\ref{average rate}.
{\color{black}
In this figure, although the normalized average rates ($R_1$ and $R_2$) decrease as $m$ increases,
the  overall average rates of the users ($mR_1$ and $mR_2$) still increase with $m$.}
}

\begin{figure}
\begin{minipage}[tbp]{0.5\linewidth}
\centering
	\includegraphics[width=2.2in]{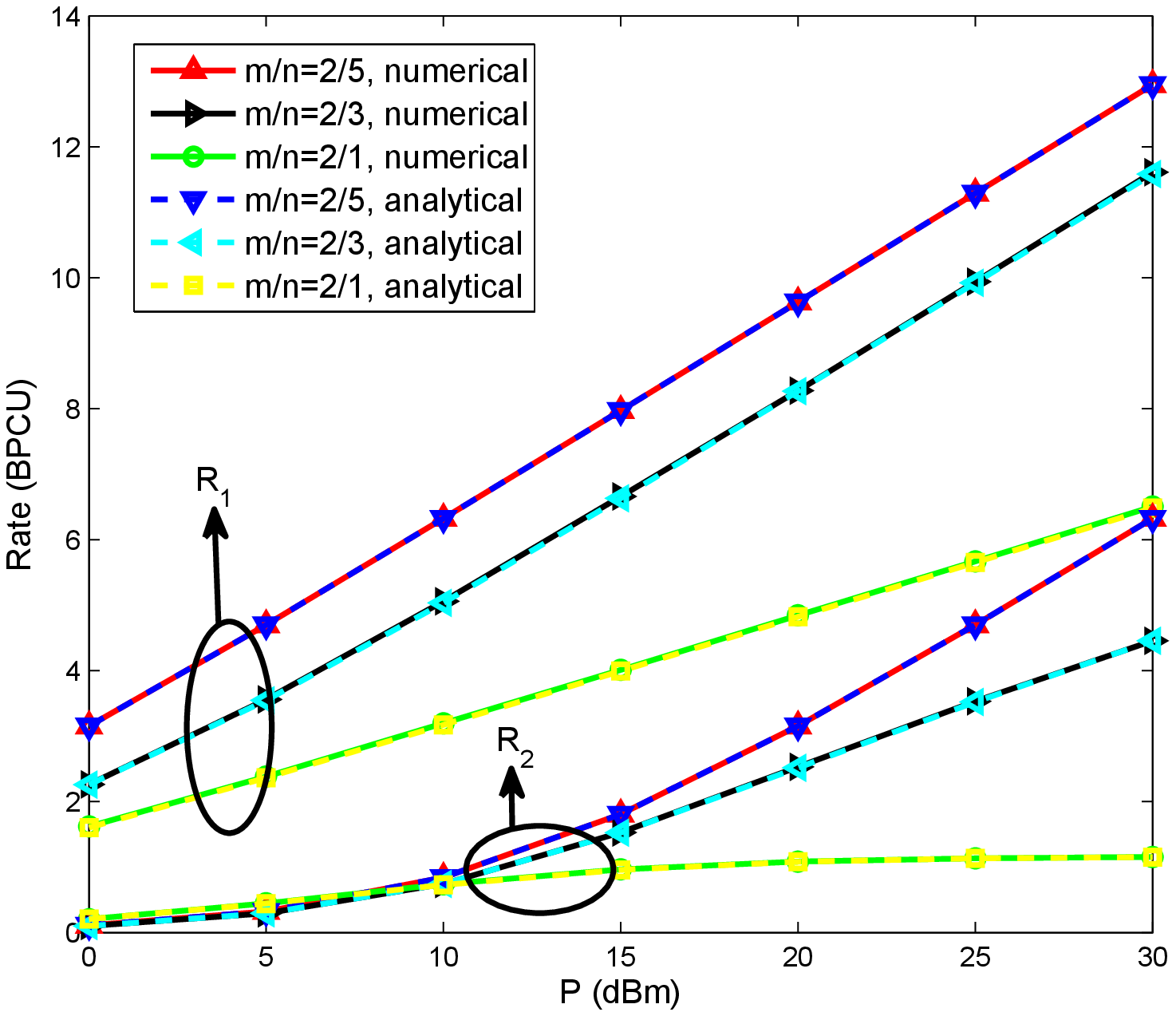}
\vspace{-2em}
    \caption{{\color{black}The normalized average individual rates of the two users versus transmission power.}}
    \label{Rate12}
    \vspace{-1em}
\end{minipage}
\begin{minipage}[tbp]{0.5\linewidth}
\centering
	\includegraphics[width=2.2in]{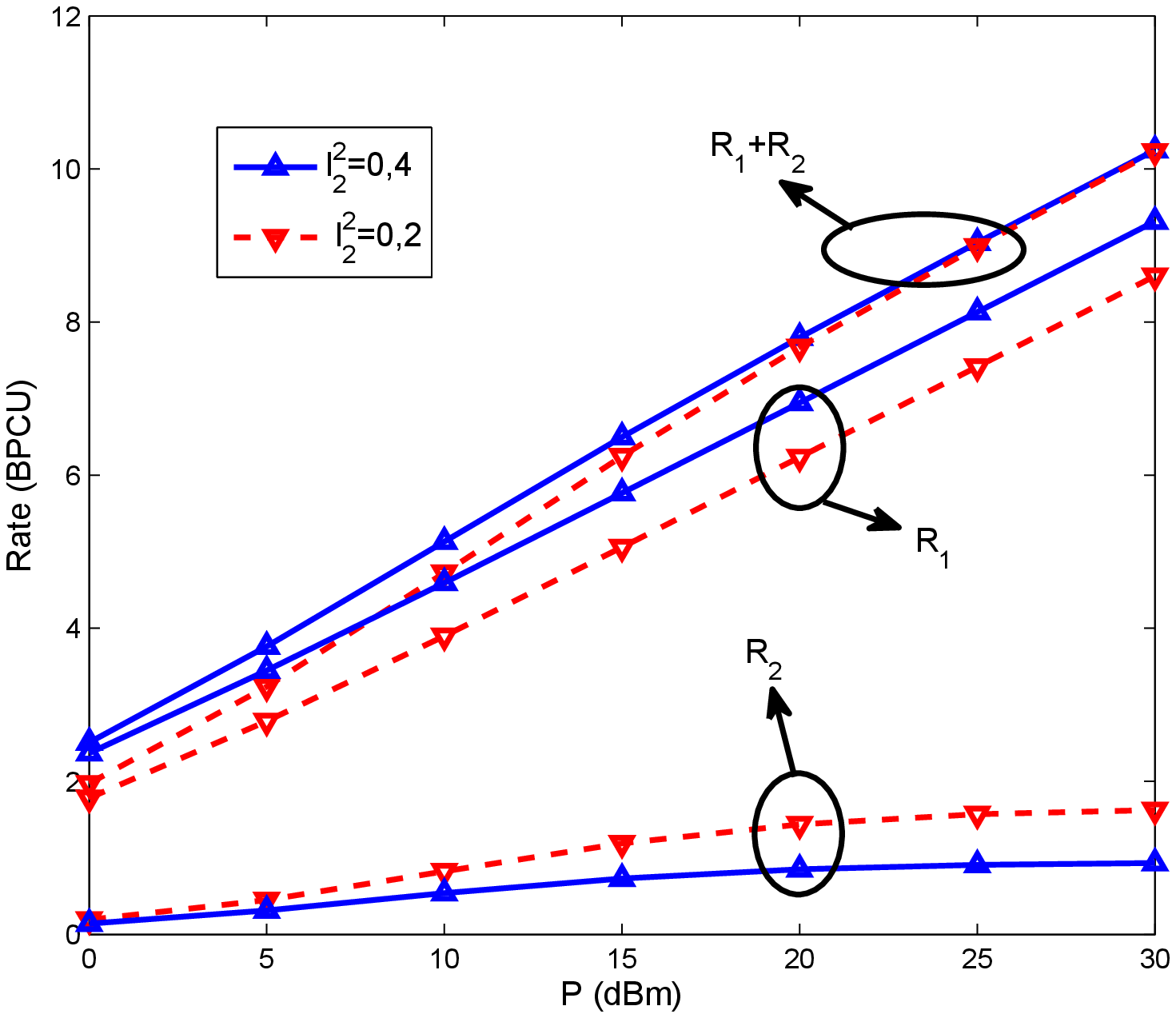}
\vspace{-2em}
    \caption{The normalized average rates of the two users for different power allocation coefficients.}
    \label{l2}
       \vspace{-1em}
\end{minipage}
\end{figure}

{\color{black}
Fig. \ref{Rate12} shows  the normalized   average individual rates of the two users
when  $m=2$, $d_1=10$m, $d_2=100$m, $\tau=2$,  $N_0=-35$dBm, and $l_2^2=0.2$.}
From this figure, we observe that the
numerical results match
the analytical results perfectly and
 the normalized  average individual rates of both users
increase when the number of  transmit antennas increases,  i.e., increasing the  number of  transmit antennas
can improve the performance of the proposed GSVD-NOMA scheme.
Moreover, as expected, user 1 which is closer to the base station  achieves a higher average rate than user 2.

{\color{black}
Please note that
in the considered asymptotic scenario,
when {\color{black} both users have the same distance to the base station, i.e,
$d_1=d_2$,
for the proposed GSVD-NOMA scheme,
the normalized average individual rates
of both users  are the same, i.e., $R_1 = R_2$, as shown in Fig. \ref{new1}.
Hence,
the proposed GSVD-NOMA scheme ensures user fairness.
When $d_1 \neq d_2$, there can be significant difference
between the users' channel conditions. In this scenario,
the use of the proposed GSVD-NOMA scheme can efficiently
explore the dynamic range of channel difference to improve
the overall system throughput. However,
we note that,  when $d_1 \neq d_2$,
the near user will achieve a higher information rate as shown in Fig. \ref{R1R2} and Fig. \ref{Rate12}, i.e.,
the overall system throughput
is improved at a price of reduced user fairness.
This is reasonable,}
since a
 user with better channel conditions
can support a higher
achievable rate,
and allocating more bandwidth resources
to the weak user is not beneficial
to the spectral efficiency.
}

Fig. \ref{l2} shows the
impact of different power allocation coefficients on GSVD-NOMA
when $m=7$, $n=5$, $d_1=10$m, $d_2=100$m, $\tau=2$, and $N_0=-35$dBm.
{\color{black}
From this figure,
we observe that
when the power allocation coefficient $l_2^2$
increases,  the normalized average rate $R_1$ of user $1$
increases and the normalized average rate $R_2$ of  user $2$
decreases, since
more powers have been allocated to
the near user.} {\color{black}On the other hand, for large $P$, i.e., in the high SNR regime, the normalized
average sum rate  of the two users $R_1+R_2$
remains almost the same when the  power allocation coefficient $l_2^2$ changes from
$0.2$ to $0.4$. Thus, in the high SNR regime,
$l_2^2$  has an insignificant impact on the average sum rate.}

\begin{figure}
\begin{minipage}[tbp]{0.5\linewidth}
\centering
	\includegraphics[width=2.2in]{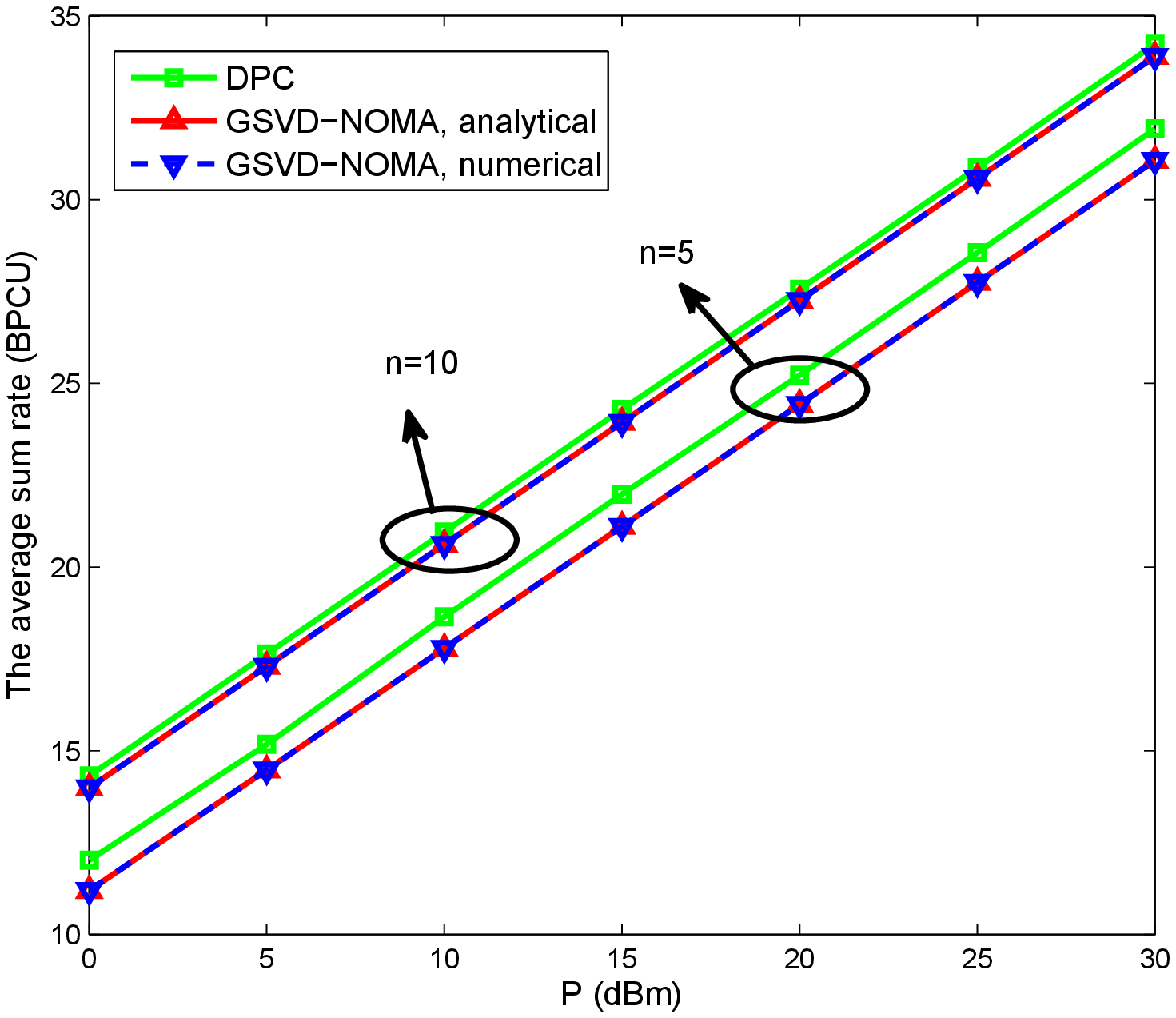}
\vspace{-2em}
    \caption{The average sum rates achieved by DPC and GSVD-NOMA.}
    \label{DPCFIG}
    \vspace{-1em}
\end{minipage}
\begin{minipage}[tbp]{0.5\linewidth}
\centering
	\includegraphics[width=2.2in]{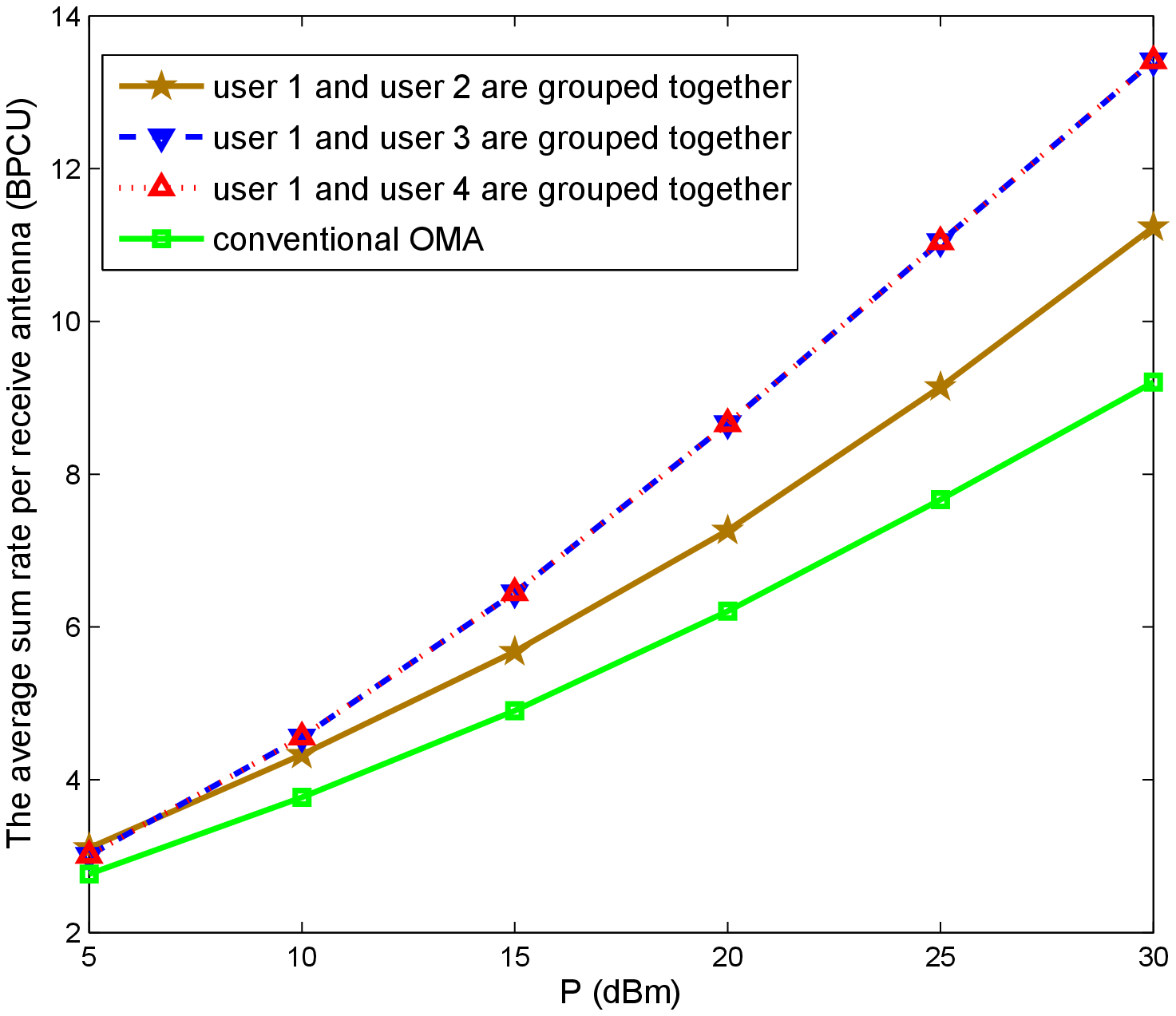}
\vspace{-2em}
    \caption{The impact of user clustering on the hybrid GSVD-NOMA system}
         \vspace{-1em}
    \label{rep1}
\end{minipage}
\end{figure}

{\color{black}
Fig. \ref{DPCFIG} compares  the   average sum rates achieved by  DPC and GSVD-NOMA
in a downlink massive MIMO scenario with one base station and two users, where the base station is equipped
with $n$ antennas and each user is equipped
with single antenna.
Moreover, it is assumed that
$d_1=10$m, $d_2=10$m, $\tau=2$, and  $N_0=-35$dBm.
From this figure, we observe that the
numerical results match
the analytical results perfectly and
 the   average sum rate of  GSVD-NOMA
approaches  that of DPC, which
 demonstrates the performance of the
proposed GSVD-NOMA scheme, and verifies the correctness of the developed analytical
results.}

{\color{black}

Although this paper focuses on the special case of two NOMA users,
the proposed GSVD-NOMA scheme
can be easily extended to
the MIMO
downlink communication scenario with
more than two users
 by
using a  hybrid MA approach.
In particular, consider a downlink
communication scenario
with one base station and multiple users.
The base station first
divides all users
into multiple groups,  where each group contains two users.
Then, the proposed GSVD-NOMA scheme
can be implemented
within each group and different groups are allocated with
orthogonal bandwidth resources.
This hybrid NOMA scheme
can be used to e.g. upgrade
an existing OMA system
without changing its fundamental resource blocks, while realizing the  performance gains of GSVD-NOMA.
The reason for   including only
two users in each group
is
to reduce the processing complexity of the SIC receiver,
and the two-user pairing consideration is consistent with the
specifications in 3GPP-LTE system \cite{As10}.}

{\color{black}

In order to demonstrate
the performance of hybrid GSVD-NOMA, we
consider a multiuser scenario with one base station and four users.
The distance between user 1 (user 2)
and the base station is
$d_1= 15$m ($d_2= 10$m), and
the distance between user 3 (user 4)
and the base station is
$d_3= 200$m ($d_4= 300$m).
In Fig. \ref{rep1}, we compare the normalized average sum rates achieved by OMA and
 hybrid GSVD-NOMA
for $m=40$, $n=50$, $\tau=2$, $N_0=-35$dBm, and $l_2^2=0.2$.
From the figure, one can observe that
no matter how the users are grouped,
the proposed  GSVD-NOMA scheme always achieves a larger
normalized average sum rate than OMA.
Moreover,
pairing user 1 with user 2 results in  the worst performance among the three possible pairing strategies.
{\color{black} This suggests that
it is preferable to pair
 users which are close to the base station
with users which are far from the base station\footnote{{\color{black}
Obviously
the performance of this
hybrid NOMA scheme
depends on
how  the users are grouped, where sophisticated algorithms
such as monotonic optimization, machine learning, and game theory can be used
for performance optimization \cite{sun2017optimal,C1,C2}.
We also suggest this as a topic for future work.
}}.
In addition, the performance gap between NOMA and OMA grows
as the transmit power $P$ increases.}}

\section{Conclusion} \label{se6}
In this paper,  a MIMO downlink communication scenario with one base
station and two users has been considered, where each user is equipped with $m$
antennas and the  base
station is  equipped with $n$
antennas. To fully exploit the available spatial degrees of freedom,
a transmission protocol which combines GSVD and NOMA has been proposed.
The  performance of the proposed protocols
has been characterized in the asymptotic regime,
{\color{black} where the numbers of
transmit and receive antennas approach infinity.}
To be more specific,
we have characterized {\color{black}the limiting distribution} of
the squared generalized singular values
of the two users' channel matrices.
Then, the normalized average individual rates of the two users have been analyzed.
{\color{black}
The proposed GSVD-NOMA scheme uses  GSVD
to decompose the
two-user MIMO broadcast channel into broadcast two-user SISO channels,
without interference among different  transmit antennas.
Moreover, the provided
numerical
results have shown that  GSVD-NOMA  achieves considerable improvements
in terms of the normalized average sum rate compared to conventional OMA and  QR-NOMA
in \cite{ding2016mimo},
and
when the base station and the users have moderate numbers of  antennas (e.g. $m=2, n=5$),
the derived analytical results  still provide good approximations.}
{\color{black}
Furthermore,
we use a hybrid MA approach
to
extend
the proposed GSVD-NOMA scheme
to the MIMO
downlink communication scenario with
more than two users,
where the base station first
divides
the users into different groups, then the proposed GSVD-NOMA scheme is
implemented within each group,
and different groups are allocated with
orthogonal bandwidth resources.}

   \vspace{-1em}

\section*{Appendix A: Proof of Theorem \ref{GSVDvalue}}
{\color{black}
Since the elements of $\mathbf{H}_1$ and $\mathbf{H}_2$
are i.i.d. complex Gaussian random variables, they are full rank with probability one
\cite{As09}.
Thus,
in this appendix,
it is assumed that
$\mathbf{H}_1$ and $\mathbf{H}_2$ are full rank.}
\setcounter{subsection}{0}

\subsection{The case when  $m \geq n$}
First, we revisit Zha's method \cite{zhaGSVD} to construct the
GSVD of the two $m \times n$ channel matrices $\mathbf{H}_1$ and $\mathbf{H}_2$.
Then,
we
obtain {\color{black}the limiting distribution} of {\color{black} the squared generalized singular values}
based on this GSVD  construction method.

\subsubsection{Steps of GSVD when  $m \geq n$}
{\color{black}
When $m \geq n$, we can construct the
GSVD of the two $m \times n$ channel matrices $\mathbf{H}_1$ and $\mathbf{H}_2$
with the following  steps:
\begin{itemize}
\item{Denote the SVD of matrix  $\mathbf{H}_2$ by
\begin{eqnarray}
  \mathbf{U}_2\mathbf{H}_2\mathbf{V}_2=
\left(\begin{array}{c}
\mathbf{O}_{(m-n) \times n}\\
\mathbf{\Sigma}_{H2}
\end{array}\right),\end{eqnarray}
where $\mathbf{\Sigma}_{H2}=\diag(z_1, z_2 ,\cdots  z_n)$ and $z_1\geq z_2 \cdots \geq z_n$.
Moreover, define $\mathbf{Q}_1$ as follows:
\begin{eqnarray}\mathbf{Q}_1=\mathbf{V}_2 \diag\left(\frac{1}{z_1}, \cdots, \frac{1}{z_n} \right).\end{eqnarray}
Then, define $\mathbf{H}_1'$  and $\mathbf{H}_2'$ as follows:
\begin{eqnarray}
\mathbf{H}_1'=\mathbf{H}_1\mathbf{Q}_1=\mathbf{H}_1\mathbf{V}_2 \diag\left(\frac{1}{z_1}, \cdots, \frac{1}{z_n} \right)
\quad \text{and} \quad \mathbf{H}_2'=\mathbf{U}_2\mathbf{H}_2\mathbf{Q}_1=\left(\begin{array}{c}
\mathbf{O}_{(m-n) \times n}\\
\mathbf{I}_{n}
\end{array}\right).\end{eqnarray}
}

\item{Next, define the
SVD of  matrix $\mathbf{H}_1'$ as
\begin{eqnarray} \label{ap1}\mathbf{U}_3\mathbf{H}_1'\mathbf{V}_3=
\left(\begin{array}{c}
\diag(w_1, \cdots, w_n )\\
\mathbf{O}_{(m-n) \times n}
\end{array}\right),
\end{eqnarray}
where $w_1\geq w_2 \cdots \geq w_n$ are the $n$ ordered singular values of $\mathbf{H}_1'$.
Furthermore,  define $\mathbf{Q}_2$ as follows:
\begin{eqnarray}\mathbf{Q}_2=\mathbf{V}_3 \diag(\beta_1,\beta_2, \cdots\beta_n ),\end{eqnarray}
where $\beta_i=(1+w_i^2)^{-1/2}$.
}

\item{Finally, it can be shown  that
\begin{eqnarray}\mathbf{U}_3\mathbf{H}_1'\mathbf{Q}_2=\mathbf{U}_3\mathbf{H}_1\mathbf{Q}_1\mathbf{Q}_2=
\left(\begin{array}{c}
\diag(\alpha_1, \cdots, \alpha_n )\\
\mathbf{O}_{(m-n) \times n}
\end{array}\right),\end{eqnarray}
where $\alpha_i=w_i\beta_i=\frac{w_i}{(1+w_i^2)^{1/2}}$ and it is easy to see that $\alpha_i^2+\beta_i^2=1$.
Moreover, it can be shown  that
\begin{eqnarray}
\left(\begin{array}{cc}
\mathbf{I}_{m-n} &\mathbf{O}_{(m-n) \times n}\\
\mathbf{O}_{n \times (m-n)}& \mathbf{V}_3^H
\end{array}\right) \mathbf{H}_2' \mathbf{Q}_2
&=&
\left(\begin{array}{cc}
\mathbf{I}_{m-n} &\mathbf{O}_{(m-n) \times n}\\
\mathbf{O}_{n \times (m-n)}& \mathbf{V}_3^H
\end{array}\right) \mathbf{U}_2\mathbf{H}_2\mathbf{Q}_1 \mathbf{Q}_2 \nonumber
\\
&=&\left(\begin{array}{c}
\mathbf{O}_{(m-n) \times n}\\
\diag(\beta_1, \cdots, \beta_n )
\end{array}\right) .\end{eqnarray}
Then, for the case of $m \geq n$,  we have provided the steps needed for constructing the
GSVD defined as in \eqref{GSVDFORM} with $\mathbf{Q}=\mathbf{Q}_1 \mathbf{Q}_2$, $\mathbf{U}=\mathbf{U}_3$ and
$\mathbf{V}=\left(\begin{array}{cc}
\mathbf{I}_{m-n} &\mathbf{O}_{(m-n) \times n}\\
\mathbf{O}_{n \times (m-n)}& \mathbf{V}_3^H
\end{array}\right) \mathbf{U}_2$.}

\end{itemize}

}

\subsubsection{{\color{black}The limiting distribution} of the squared generalized singular values when  $m \geq n$}
\label{cz}
From \eqref{ap1}, it can be  shown that
$ \mathbf{U}_3\mathbf{H}_1'\mathbf{V}_3=
\left(\begin{array}{c}
\diag(w_1, \cdots, w_n )\\
\mathbf{O}_{(m-n) \times n}
\end{array}\right),
$
and
{\color{black} the squared generalized singular values} $w_i^2=\alpha_i^2/\beta_i^2$, $i \in \{ 1, \cdots, n \}$,
are the eigenvalues of $\mathbf{H}_1'^H\mathbf{H}_1'$, where
$\mathbf{H}_1'=\mathbf{H}_1\mathbf{Q}_1=\mathbf{H}_1\mathbf{V}_2 \diag(\frac{1}{z_1}, \cdots, \frac{1}{z_n} )$.
Recall that the SVD of  $\mathbf{H}_2$ is $ \mathbf{U}_2\mathbf{H}_2\mathbf{V}_2=
\left(\begin{array}{c}
\mathbf{O}_{(m-n) \times n}\\
\diag(z_1,\cdots  z_n)
\end{array}\right)
$, and we have that $\mathbf{V}_2^H \mathbf{H}_2^H\mathbf{H}_2\mathbf{V}_2=\diag(z_1, \cdots, z_n)^2$.
Also, since
$\mathbf{H}_2^H\mathbf{H}_2$ is a central Wishart matrix which is unitarily  invariant \cite{tulino2004foundations},
it can be shown that $\mathbf{V}_2$ is a Haar matrix which is independent of $\diag(z_1, \cdots, z_n )^2$.
So $\mathbf{H}_1'=\mathbf{H}_1\mathbf{V}_2 \diag(\frac{1}{z_1}, \cdots, \frac{1}{z_n} )$ is the product of
three independent matrices
$\mathbf{H}_1$, $\mathbf{V}_2$, and  $\diag(\frac{1}{z_1}, \cdots, \frac{1}{z_n} )$.

Define $\mathbf{G}_1=\mathbf{H}_1\mathbf{V}_2$. Then,
it can be shown that $\mathbf{G}_1$ is an $m \times n$ matrix whose elements are i.i.d. complex Gaussian random variables with
zero mean and unit variance. Furthermore, we have that
$\mathbf{H}_1'=\mathbf{G}_1\diag(\frac{1}{z_1}, \cdots, \frac{1}{z_n} )$ and
{\color{black} the squared generalized singular values} $w_i^2$, $i \in \{1, \cdots,n\}$, can be expressed as the eigenvalues of
$\mathbf{H}_1'^H\mathbf{H}_1'=\diag(\frac{1}{z_1}, \cdots,
\frac{1}{z_n} )\mathbf{G}_1^H\mathbf{G}_1\diag(\frac{1}{z_1}, \cdots, \frac{1}{z_n} )$.
Moreover, {\color{black} the squared generalized singular values} $w_i^2$, $i \in \{1, \cdots,n\}$, can be also expressed as the eigenvalues of
$\mathbf{G}_1^H\mathbf{G}_1\diag(\frac{1}{z_1}, \cdots, \frac{1}{z_n} )^2$
or the nonzero eigenvalues of
$\mathbf{G}_1\diag(\frac{1}{z_1}, \cdots, \frac{1}{z_n} )^2\mathbf{G}_1^H$.
{\color{black}
Note that $\mathbf{G}_1$ is independent of $z_i$, $i \in \{1,\cdots, n\}$,
where $z_i$ is a nonzero singular value of $\mathbf{H}_2$.
Thus, directly from \cite[Theorem 2.39 and Theorem 2.40]{tulino2004foundations},
the e.d.f. of the nonzero eigenvalues  of $\mathbf{G}_1\diag(\frac{1}{z_1}, \cdots, \frac{1}{z_n} )^2\mathbf{G}_1^H$,
which is identical to the e.d.f. of the generalized singular values $\mathbf{F}_{w_i}^n(x)$, converges, almost surely,
as $m,n \to \infty$ with $\frac{m}{n}\to \eta$,
to
a nonrandom c.d.f. $\mathbf{F}_{w_i}(x)$, whose p.d.f. is $f_{\frac{1}{\eta}, \frac{1}{\eta}}(x)$.}
Then, when $m \geq n$, we have characterized the limiting distribution of  the squared  generalized singular values $w_i^2$.

   \vspace{-1em}
\subsection{The case when  $m<n<2m$}
Again, we first use Zha's method  to construct the
GSVD of the two $m \times n$ channel matrices $\mathbf{H}_1$ and $\mathbf{H}_2$.

\subsubsection{Steps of GSVD when  $m<n<2m$}
{\color{black}
When $m<n<2m$, we can construct the
GSVD of the two $m \times n$ channel matrices $\mathbf{H}_1$ and $\mathbf{H}_2$
with the following  steps:
\begin{itemize}
\item{First, we define
the SVD of matrix  $\mathbf{H}_2$ as \begin{eqnarray} \mathbf{U}_2\mathbf{H}_2\mathbf{V}_2=
\left(\begin{array}{cc}
\mathbf{O}_{m \times (n-m)} &\diag(z_1, z_2 ,\cdots  z_m)
\end{array}\right),
\end{eqnarray} where $z_1\geq z_2 \cdots \geq z_m$.
Let us define $\mathbf{Q}_1$ as follows:
\begin{eqnarray}
\mathbf{Q}_1=\mathbf{V}_2
\left(\begin{array}{cc}
\mathbf{I}_{n-m} & \mathbf{O}_{(n-m) \times m}\\
\mathbf{O}_{m \times (n-m)} & \diag(z_1, \cdots  z_m)^{-1}
\end{array}\right).\end{eqnarray}
Then, it can be shown that \begin{eqnarray}\mathbf{H}_1'=\mathbf{H}_1\mathbf{Q}_1=\mathbf{H}_1\mathbf{V}_2 \left(\begin{array}{cc}
\mathbf{I}_{n-m} & \mathbf{O}_{(n-m) \times m}\\
\mathbf{O}_{m \times (n-m)} & \diag(z_1, \cdots  z_m)^{-1}
\end{array}\right)=
\left(\begin{array}{cc}
\mathbf{H}_{11}' & \mathbf{H}_{12}'
\end{array}\right),
\end{eqnarray} where  $\mathbf{H}_{11}'=\mathbf{H}_1\mathbf{V}_2
\left(\begin{array}{c}
\mathbf{I}_{n-m}\\
\mathbf{O}_{m \times (n-m)}
\end{array}\right) $ and $\mathbf{H}_{12}'=\mathbf{H}_1\mathbf{V}_2
\left(\begin{array}{c}
\mathbf{O}_{(n-m) \times m}\\
\diag(z_1, \cdots  z_m)^{-1}
\end{array}\right)$. Also, $\mathbf{H}_2'$ can be defined as follows:
\begin{eqnarray}
\mathbf{H}_2'&=&\mathbf{U}_2\mathbf{H}_2\mathbf{Q}_1=
\left(\begin{array}{cc}
\mathbf{O}_{m \times (n-m)}  &
\diag(z_1, \cdots  z_m)
\end{array}\right)
\left(\begin{array}{cc}
\mathbf{I}_{n-m} & \mathbf{O}_{(n-m) \times m}\\
\mathbf{O}_{m \times (n-m)} & \diag(z_1, \cdots  z_m)^{-1}
\end{array}\right) \nonumber
\\
&=&\left(\begin{array}{cc}
\mathbf{O}_{m \times (n-m)} &
\mathbf{I}_m
\end{array}\right).
\end{eqnarray}
}

\item{It is easy to see that the rank of $\mathbf{H}_{11}'$ is $r$, where $r=n-m$.
Then, we can define the
SVD of  matrix $\mathbf{H}_{11}'$ as follows:
\begin{eqnarray} \mathbf{U}_{11}\mathbf{H}_{11}'\mathbf{V}_{11}=
\left(\begin{array}{c}
\mathbf{\Sigma}_A\\
\mathbf{O}_{(m-r) \times r}
\end{array}\right),
\end{eqnarray}
where $\mathbf{\Sigma}_{A}=\diag(t_1, \cdots, t_r)$
with $t_1\geq t_2 \cdots \geq t_r > 0$.
Furthermore,  let us define $\mathbf{Q}_2$ as follows:
\begin{eqnarray}
\mathbf{Q}_2=
\left(\begin{array}{cc}
\mathbf{V}_{11} &\mathbf{O}_{r \times m}\\
\mathbf{O}_{m \times r}& \mathbf{I}_m
\end{array}\right)
\left(\begin{array}{cc}
\mathbf{\Sigma}_{A}^{-1} &\mathbf{O}_{r \times m}\\
\mathbf{O}_{m \times r}& \mathbf{I}_{m}
\end{array}\right).\end{eqnarray} Then, it can be shown that
\begin{eqnarray}
\mathbf{H}_2''=\mathbf{H}_2'\mathbf{Q}_2
=\left(\begin{array}{cc}
\mathbf{O}_{m \times (n-m)} &
\mathbf{I}_m
\end{array}\right)
\mathbf{Q}_2= \left(\begin{array}{cc}
\mathbf{O}_{m \times (n-m)} &
\mathbf{I}_m
\end{array}\right), \quad \text{and}\end{eqnarray}
\begin{eqnarray}
\mathbf{H}_1''=\mathbf{U}_{11} \mathbf{H}_1' \mathbf{Q}_2
&=& \nonumber
\left(\begin{array}{cc}
\mathbf{U}_{11}\mathbf{H}_{11}' & \mathbf{U}_{11}\mathbf{H}_{12}'
\end{array}\right)
\left(\begin{array}{cc}
\mathbf{V}_{11} &\mathbf{O}_{r \times m}\\
\mathbf{O}_{m \times r}& \mathbf{I}_m
\end{array}\right)
\left(\begin{array}{cc}
\mathbf{\Sigma}_{A}^{-1} &\mathbf{O}_{r \times m}\\
\mathbf{O}_{m \times r}& \mathbf{I}_{m}
\end{array}\right)
\\ \nonumber
&=&\left(\begin{array}{cc}
\left(\begin{array}{c}
\mathbf{\Sigma}_A\\
\mathbf{O}_{(m-r) \times r}
\end{array}\right)& \mathbf{U}_{11}\mathbf{H}_{12}'
\end{array}\right)\left(\begin{array}{cc}
\mathbf{\Sigma}_{A}^{-1} &\mathbf{O}_{r \times m}\\
\mathbf{O}_{m \times r}& \mathbf{I}_{m}
\end{array}\right)
\\
&=&\left(\begin{array}{cc}
\left(\begin{array}{c}
\mathbf{I}_r\\
\mathbf{O}_{(m-r) \times r}
\end{array}\right)& \mathbf{U}_{11}\mathbf{H}_{12}'
\end{array}\right)
=
\left(\begin{array}{cc}
\mathbf{I}_{r} & \mathbf{A}_{13}\\
\mathbf{O}_{(m-r) \times r} & \mathbf{A}_{23}
\end{array}\right),
\end{eqnarray}

where $\left(\begin{array}{c}
 \mathbf{A}_{13}\\
\mathbf{A}_{23}
\end{array}\right)=\mathbf{U}_{11}\mathbf{H}_{12}'$}, $\mathbf{A}_{13} \in \mathbb{C}^{r \times (n-r)}$, and $\mathbf{A}_{23}\in \mathbb{C}^{(m-r) \times (n-r)}$.

\item{It is
easy to see that the rank of $\mathbf{A}_{23}$ is $m-r$.
Let us define {\color{black} $q=2m-n=m-r$}.
Then, we can rewrite the
SVD of matrix $\mathbf{A}_{23}$ as
\begin{eqnarray} \mathbf{U}_{23}\mathbf{A}_{23}\mathbf{V}_{23}=
\left(\begin{array}{cc}
\label{ap2}
\mathbf{\Sigma}_{A_{23}} &
\mathbf{O}_{q \times r}
\end{array}\right),
\end{eqnarray}
where $\mathbf{\Sigma}_{A_{23}}=\diag(w_1, \cdots, w_q)$
with $w_1\geq w_2 \cdots \geq w_q > 0$.
Also, let us define
$\mathbf{S}_{2}=\diag(\beta_1, \cdots, \beta_q)$
with $\beta_i=(1+w_i^2)^{-1/2}$
and $\mathbf{Q}_3$ as follows:
\begin{eqnarray}
\mathbf{Q}_3=\left(\begin{array}{cc}
\mathbf{I}_{r} & - \mathbf{A}_{13}\\
\mathbf{O}_{m \times r} & \mathbf{I}_{m}
\end{array}\right)
\left(\begin{array}{cc}
\mathbf{I}_{r} & \mathbf{O}_{r \times m}\\
\mathbf{O}_{m \times r} & \mathbf{V}_{23}
\end{array}\right)
\diag(
\mathbf{I}_{r} , \mathbf{S}_{2} , \mathbf{I}_{r}
).
\end{eqnarray}
}

\item{Finally, it can be shown that
\begin{eqnarray}
&&\mathbf{V}_{23}^H\mathbf{H}_2''\mathbf{Q}_3=\mathbf{V}_{23}^H \mathbf{U}_2 \mathbf{H}_2 \mathbf{Q}_1 \mathbf{Q}_2 \mathbf{Q}_3
=
\mathbf{V}_{23}^H
\left(\begin{array}{cc}
\mathbf{O}_{m \times r} &
\mathbf{I}_m
\end{array}\right)
\left(\begin{array}{cc}
\mathbf{I}_{r} & - \mathbf{A}_{13}\\
\mathbf{O}_{m \times r} & \mathbf{I}_{m}
\end{array}\right)
\\ \nonumber&& \times
\left(\begin{array}{cc}
\mathbf{I}_{r} & \mathbf{O}_{r \times m}\\
\mathbf{O}_{m \times r} & \mathbf{V}_{23}
\end{array}\right)
\diag(
\mathbf{I}_{r} , \mathbf{S}_{2} , \mathbf{I}_{r}
)
=\left(\begin{array}{ccc}
\mathbf{O}_{q \times r}&\mathbf{S}_2&\mathbf{O}_{q \times r}\\
\mathbf{O}_{r \times r}&\mathbf{O}_{r \times q}&\mathbf{I}_r
\end{array}\right).
\end{eqnarray}

Moreover, it can be shown that
\begin{eqnarray}
&&\left(\begin{array}{cc}
\mathbf{I}_r &\mathbf{O}_{r \times q}\\
\mathbf{O}_{q \times r}& \mathbf{U}_{23}
\end{array}\right) \mathbf{H}_1''\mathbf{Q}_3=\left(\begin{array}{cc}
\mathbf{I}_r &\mathbf{O}_{r \times q}\\
\mathbf{O}_{q \times r}& \mathbf{U}_{23}
\end{array}\right)  \mathbf{U}_{11}   \mathbf{H}_1 \mathbf{Q}_1 \mathbf{Q}_2 \mathbf{Q}_3
=\left(\begin{array}{cc}
\mathbf{I}_r &\mathbf{O}_{r \times q}\\
\mathbf{O}_{q \times r}& \mathbf{U}_{23}
\end{array}\right) \nonumber
\\ \nonumber&& \times
\left(\begin{array}{cc}
\mathbf{I}_{r} & \mathbf{A}_{13}\\
\mathbf{O}_{q \times r} & \mathbf{A}_{23}
\end{array}\right)\left(\begin{array}{cc}
\mathbf{I}_{r} & - \mathbf{A}_{13}\\
\mathbf{O}_{m \times r} & \mathbf{I}_{m}
\end{array}\right)
\left(\begin{array}{cc}
\mathbf{I}_{r} & \mathbf{O}_{r \times m}\\
\mathbf{O}_{m \times r} & \mathbf{V}_{23}
\end{array}\right)
\diag(
\mathbf{I}_{r} , \mathbf{S}_{2} , \mathbf{I}_{r}
)=
\\  &&
\left(\begin{array}{ccc}
\mathbf{I}_r&\mathbf{O}_{r \times q} & \mathbf{O}_{r \times r}\\
\mathbf{O}_{q \times r}&\mathbf{\Sigma}_{A_{23}} \mathbf{S}_{2}&\mathbf{O}_{q \times r}
\end{array}\right).
\end{eqnarray}
Furthermore, define $\mathbf{S}_{1}=\mathbf{\Sigma}_{A_{23}} \mathbf{S}_{2}=\diag(\alpha_1, \cdots, \alpha_q)$
with $\alpha_i=w_i\beta_i=\frac{w_i}{(1+w_i^2)^{1/2}}$.
Then, for the case  $m<n<2m$,  we have provided the steps for constructing the
GSVD defined in \eqref{GSVDFORM} with $\mathbf{Q}=\mathbf{Q}_1 \mathbf{Q}_2 \mathbf{Q}_3$, $\mathbf{U}=\left(\begin{array}{cc}
\mathbf{I}_r &\mathbf{O}_{r \times q}\\
\mathbf{O}_{q \times r}& \mathbf{U}_{23}
\end{array}\right)  \mathbf{U}_{11} $, and
$\mathbf{V}=\mathbf{V}_{23}^H \mathbf{U}_2$.
}
\end{itemize}}

\subsubsection{{\color{black}The limiting distribution} of the squared generalized singular values when  $m<n<2m$}
From \eqref{ap2}, we can see that
$ \mathbf{U}_{23}\mathbf{A}_{23}\mathbf{V}_{23}=
\left(\begin{array}{cc}
\diag(w_1, \cdots, w_{2m-n}) &
\mathbf{O}_{(2m-n) \times (n-m)}
\end{array}\right),
$
and
{\color{black} the squared generalized singular values} $w_i^2=\alpha_i^2/\beta_i^2$, $i \in \{1, \cdots, 2m-n\}$,
are the eigenvalues of $\mathbf{A}_{23}\mathbf{A}_{23}^H$.
Recall that
\begin{eqnarray}
\left(\begin{array}{c}
 \mathbf{A}_{13}\\
\mathbf{A}_{23}
\end{array}\right)=\mathbf{U}_{11}\mathbf{H}_{12}',
\quad
\mathbf{H}_1\mathbf{V}_2 \left(\begin{array}{cc}
\mathbf{I}_{n-m} & \mathbf{O}_{(n-m) \times m}\\
\mathbf{O}_{m \times (n-m)} & \diag(z_1, \cdots,  z_m)^{-1}
\end{array}\right)=
\left(\begin{array}{cc}
\mathbf{H}_{11}' & \mathbf{H}_{12}'
\end{array}\right),\end{eqnarray}
$ \mathbf{U}_2\mathbf{H}_2\mathbf{V}_2=
\left(\begin{array}{cc}
\mathbf{O}_{m \times (n-m)} &\diag(z_1, \cdots,  z_m)
\end{array}\right),
$
and
$ \mathbf{U}_{11}\mathbf{H}_{11}'\mathbf{V}_{11}=
\left(\begin{array}{c}
\mathbf{\Sigma}_A\\
\mathbf{O}_{(2m-n) \times (n-m)}
\end{array}\right).
$
Similarly, from \cite{tulino2004foundations}, it can be shown that
$\mathbf{V}_2$ is a Haar matrix which is independent of $\diag(z_1, \cdots,  z_m)$.
Let us define $\mathbf{P}=\mathbf{H}_1\mathbf{V}_2=\left( \mathbf{P}_1 \quad \mathbf{P}_2 \right)$ with
$\mathbf{P}_1 \in \mathbb{C}^{m \times (n-m)}$ and  $\mathbf{P}_2 \in \mathbb{C}^{m \times m}$.
Then, it can be shown that $\mathbf{P}$ is an $m \times n$ matrix whose elements are i.i.d. complex Gaussian random variables with
zero mean and unit variance and independent of  $\diag(z_1, \cdots,  z_m)$.
It is easy to see that $\mathbf{P}_1=\mathbf{H}_{11}'$, so
$\mathbf{H}_{11}'$ is independent of $\mathbf{P}_2$. Thus,
$\mathbf{U}_{11}$ is independent of $\mathbf{P}_2$. Then, $\mathbf{W}=\mathbf{U}_{11}\mathbf{P}_2$ is
an $m \times m$ matrix whose elements are i.i.d. complex Gaussian random variables with
zero mean and unit variance and independent of   $\diag(z_1, \cdots,  z_m)$.

Let $\mathbf{W}^H=(\mathbf{W}_1^H \quad \mathbf{W}_2^H)$
with
$\mathbf{W}_1 \in \mathbb{C}^{(n-m) \times m}$ and  $\mathbf{W}_2 \in \mathbb{C}^{(2m-n) \times m}$.
Then, we can rewrite $\mathbf{A}_{23}$ as $\mathbf{A}_{23}=\mathbf{W}_2 \diag(z_1, \cdots,  z_m)^{-1}$.
{\color{black} The squared generalized singular values} $w_i^2$, $i \in \{1, \cdots, 2m-n\}$, can be expressed as the eigenvalues of
$\mathbf{W}_2 \diag(z_1, \cdots,  z_m) ^{-2}\mathbf{W}_2^H$.
%
%
{\color{black}
Note that $\mathbf{W}_2$ is independent of $z_i$, $i \in \{1,\cdots, m\}$,
where $z_i$ is a nonzero singular value of $\mathbf{H}_2$.
Thus, directly from \cite[Theorem 2.39 and Theorem 2.40]{tulino2004foundations},
the e.d.f. of the nonzero eigenvalues  of $\mathbf{W}_2 \diag(z_1, \cdots,  z_m) ^{-2}\mathbf{W}_2^H$,
which is identical to the e.d.f. of the generalized singular values $\mathbf{F}_{w_i}^{2m-n}(x)$,
converges, almost surely,
as $m,n \to \infty$ with $\frac{m}{n}\to \eta$,
to a nonrandom c.d.f. $\mathbf{F}_{w_i}(x)$, whose
p.d.f. is  $\frac{\eta}{(2\eta-1)^2}f_{\frac{\eta}{2\eta-1}, \eta}\left(\frac{x}{2\eta-1}\right).$}
Then, when $m<n<2m$, we have characterized the limiting distribution of  the squared  generalized singular values $w_i^2$.

\subsection{The case when  $2m \leq n$}
For the case of $2m \leq n$, the construction of the GSVD is different from those described before. In the following, we again use Zha's method to construct the
GSVD of the two $m \times n$ channel matrices $\mathbf{H}_1$ and $\mathbf{H}_2$.

\subsubsection{Steps of GSVD when  $2m<n$}
{\color{black}
When $2m<n$, we can construct the
GSVD of the two $m \times n$ channel matrices $\mathbf{H}_1$ and $\mathbf{H}_2$
with the following  steps:
\begin{itemize}
\item{First, we define
the SVD of matrix  $\mathbf{H}_2$ as \begin{eqnarray} \mathbf{U}_2\mathbf{H}_2\mathbf{V}_2=
\left(\begin{array}{cc}
\mathbf{O}_{m \times (n-m)} &\diag(z_1, z_2 ,\cdots  z_m)
\end{array}\right),\end{eqnarray} where $z_1\geq z_2 \cdots \geq z_m$.
Let us define $\mathbf{Q}_1=\mathbf{V}_2
\left(\begin{array}{cc}
\mathbf{I}_{n-m} & \mathbf{O}_{(n-m) \times m}\\
\mathbf{O}_{m \times (n-m)} & \diag(z_1, \cdots  z_m)^{-1}
\end{array}\right)
$.
Then, it can be shown that \begin{eqnarray}\mathbf{H}_1'=\mathbf{H}_1\mathbf{Q}_1=\mathbf{H}_1\mathbf{V}_2 \left(\begin{array}{cc}
\mathbf{I}_{n-m} & \mathbf{O}_{(n-m) \times m}\\
\mathbf{O}_{m \times (n-m)} & \diag(z_1, \cdots  z_m)^{-1}
\end{array}\right)=
\left(\begin{array}{cc}
\mathbf{H}_{11}' & \mathbf{H}_{12}'
\end{array}\right),\end{eqnarray} where  $\mathbf{H}_{11}'=\mathbf{H}_1\mathbf{V}_2
\left(\begin{array}{c}
\mathbf{I}_{n-m}\\
\mathbf{O}_{m \times (n-m)}
\end{array}\right) $, $\mathbf{H}_{12}'=\mathbf{H}_1\mathbf{V}_2
\left(\begin{array}{c}
\mathbf{O}_{(n-m) \times m}\\
\diag(z_1, \cdots  z_m)^{-1}
\end{array}\right)$,
$\mathbf{H}_{11}' \in \mathbb{C}^{m \times (n-m)}$, and $\mathbf{H}_{12}'\in \mathbb{C}^{m \times m}$. Also, it can be shown that
\begin{eqnarray}
\mathbf{H}_2'&=&\mathbf{U}_2\mathbf{H}_2\mathbf{Q}_1
=
\left(\begin{array}{cc}
\mathbf{O}_{m \times (n-m)} &
\diag(z_1, \cdots  z_m)
\end{array}\right)
\\ \nonumber&& \times
\left(\begin{array}{cc}
\mathbf{I}_{n-m} & \mathbf{O}_{(n-m) \times m}\\
\mathbf{O}_{m \times (n-m)} & \diag(z_1, \cdots  z_m)^{-1}
\end{array}\right)
=\left(\begin{array}{cc}
\mathbf{O}_{m \times (n-m)} &
\mathbf{I}_m
\end{array}\right).
\end{eqnarray}
}

\item{It is easy to see that the rank of $\mathbf{H}_{11}'$ is $m$.
Then, we can define the
SVD of  matrix $\mathbf{H}_{11}'$  as
\begin{eqnarray} \mathbf{U}_{11}\mathbf{H}_{11}'\mathbf{V}_{11}=
\left(\begin{array}{cc}
\mathbf{\Sigma}_A &  \mathbf{O}_{m \times (n-2m)}
\end{array}\right),\end{eqnarray}
where $\mathbf{\Sigma}_{A}=\diag(t_1, \cdots, t_m)$
with $t_1\geq t_2 \cdots \geq t_m > 0$.
Furthermore,  let us define $\mathbf{Q}_2=
\left(\begin{array}{cc}
\mathbf{V}_{11} &\mathbf{O}_{(n-m) \times m}\\
\mathbf{O}_{m \times (n-m)}& \mathbf{I}_m
\end{array}\right)
\left(\begin{array}{cc}
\mathbf{\Sigma}_{A}^{-1} &\mathbf{O}_{m \times (n-m)}\\
\mathbf{O}_{(n-m) \times m}& \mathbf{I}_{n-m}
\end{array}\right)
$. Then, it can be shown that
\begin{eqnarray}\mathbf{H}_2''=\mathbf{H}_2'\mathbf{Q}_2
=\left(\begin{array}{cc}
\mathbf{O}_{m \times (n-m)} &
\mathbf{I}_m
\end{array}\right)
\mathbf{Q}_2= \left(\begin{array}{cc}
\mathbf{O}_{m \times (n-m)} &
\mathbf{I}_m
\end{array}\right), \quad \text{and}\end{eqnarray}
\begin{eqnarray}
&&\mathbf{H}_1''=\mathbf{U}_{11} \mathbf{H}_1' \mathbf{Q}_2
=
\left(\begin{array}{cc}
\mathbf{U}_{11}\mathbf{H}_{11}' & \mathbf{U}_{11}\mathbf{H}_{12}'
\end{array}\right)
\left(\begin{array}{cc}
\mathbf{V}_{11} &\mathbf{O}_{(n-m) \times m}\\
\mathbf{O}_{m \times (n-m)}& \mathbf{I}_m
\end{array}\right)
\\ \nonumber&& \times
\left(\begin{array}{cc}
\mathbf{\Sigma}_{A}^{-1} &\mathbf{O}_{m \times (n-m)}\\
\mathbf{O}_{(n-m) \times m}& \mathbf{I}_{n-m}
\end{array}\right)
=\left(\begin{array}{cc}
\begin{array}{cc}
\mathbf{I}_m &
\mathbf{O}_{m \times (n-2m)}
\end{array}& \mathbf{U}_{11}\mathbf{H}_{12}'
\end{array}\right).
\end{eqnarray}
}

\item{Finally, let us define $\mathbf{Q}_3=
\left(\begin{array}{cc}
\mathbf{I}_{n-m} & \left(\begin{array}{c}-\mathbf{U}_{11}\mathbf{H}_{12}'
\\ \mathbf{O}_{(n-2m) \times m} \end{array}
\right)\\
\mathbf{O}_{m \times (n-m)}& \mathbf{I}_m
\end{array}\right)
$.
Then, it can be shown that
$\mathbf{H}_2''\mathbf{Q}_3=\mathbf{U}_2 \mathbf{H}_2 \mathbf{Q}_1 \mathbf{Q}_2 \mathbf{Q}_3
=
\left(\begin{array}{cc}
\mathbf{O}_{m \times (n-m)} &
\mathbf{I}_m
\end{array}\right)
\mathbf{Q}_3
=
\left(\begin{array}{cc}
\mathbf{O}_{m \times (n-m)} &
\mathbf{I}_m
\end{array}\right).$ Moreover,it can be shown that $\mathbf{H}_1''\mathbf{Q}_3=\mathbf{U}_{11}   \mathbf{H}_1 \mathbf{Q}_1 \mathbf{Q}_2 \mathbf{Q}_3
= \left(\begin{array}{cc}
\begin{array}{cc}
\mathbf{I}_m &
\mathbf{O}_{m \times (n-2m)}
\end{array}& \mathbf{U}_{11}\mathbf{H}_{12}'
\end{array}\right)\mathbf{Q}_3
=\left(\begin{array}{cc}
\mathbf{I}_m &
\mathbf{O}_{m \times (n-m)}
\end{array}\right).$
Thus, for  $2m<n$,  we have provided the steps for constructing the
GSVD defined in \eqref{GSVDFORM} with $\mathbf{Q}=\mathbf{Q}_1 \mathbf{Q}_2 \mathbf{Q}_3$, $\mathbf{U}=\mathbf{U}_{11} $, and
$\mathbf{V}=\mathbf{U}_2$.
}
\end{itemize}}

\subsubsection{The squared generalized singular values when  $2m<n$}
When $2m<n$, after using Zha's GSVD method, we have transformed
the two $m \times n$ channel matrices $\mathbf{H}_1$ and $\mathbf{H}_2$ into
$\left(\begin{array}{cc}
\mathbf{I}_m &
\mathbf{O}_{m \times (n-m)}
\end{array}\right)$ and $\left(\begin{array}{cc}
\mathbf{O}_{m \times (n-m)} &
\mathbf{I}_m
\end{array}\right)$, respectively. Thus, the squared generalized singular values become constants in this case.

This concludes the proof of the theorem. \hspace{\fill}$\blacksquare$\newline

\vspace{-2em}

\section*{Appendix B: Proof of Theorem  \ref{Q}}
{\color{black}
In this section, we characterize $t^2 =\mathcal{E} \{\text{trace} (\mathbf{Q}\mathbf{Q}^H)\}$ when $2m \geq n$.}
From \eqref{GSVDFORM}, it can be shown that
\begin{eqnarray}
\left(\begin{array}{cc}
\mathbf{U}
& \mathbf{O}_{m \times m}
\\
 \mathbf{O}_{m \times m} & \mathbf{V}
\end{array}\right)
\left(\begin{array}{c}
\mathbf{H}_1
\\
 \mathbf{H}_2
\end{array}\right)\mathbf{Q}=
\left(\begin{array}{c}
\mathbf{\Sigma}_1
\\
\mathbf{\Sigma}_2
\end{array}\right).
\end{eqnarray}
Moreover, let us define $\mathbf{\Sigma}=\left(\begin{array}{c}
\mathbf{\Sigma}_1
\\
\mathbf{\Sigma}_2
\end{array}\right)$. Then,
when $2m \geq n$,
from \eqref{mdayun} and \eqref{mmn1},
it can be shown that
$
\mathbf{\Sigma}^H\mathbf{\Sigma}=\left(\begin{array}{cc}
\mathbf{\Sigma}_1^H
&
\mathbf{\Sigma}_2^H
\end{array}\right) \left(\begin{array}{c}
\mathbf{\Sigma}_1
\\
\mathbf{\Sigma}_2
\end{array}\right)
= \mathbf{I}_n.$
Let us define $\mathbf{H}=\left(\begin{array}{c}
\mathbf{H}_1
\\
\mathbf{H}_2
\end{array}\right)$.
Then,
we have that
\begin{eqnarray}
\mathbf{Q}^H
\mathbf{H}^H
\left(\begin{array}{cc}
\mathbf{U}^H
& \mathbf{O}_{m \times m}
\\
 \mathbf{O}_{m \times m} & \mathbf{V}^H
\end{array}\right)
\left(\begin{array}{cc}
\mathbf{U}
& \mathbf{O}_{m \times m}
\\
 \mathbf{O}_{m \times m} & \mathbf{V}
\end{array}\right)\mathbf{H}
\mathbf{Q}=\mathbf{Q}^H
\mathbf{H}^H\mathbf{H}
\mathbf{Q}=\mathbf{\Sigma}^H\mathbf{\Sigma}=\mathbf{I}_n.
\end{eqnarray}

Thus, it can be shown that $\mathbf{H}^H\mathbf{H}=\mathbf{Q}^{-H}\mathbf{I}_n\mathbf{Q}^{-1}=\mathbf{Q}^{-H}\mathbf{Q}^{-1}$
and $\mathbf{Q}\mathbf{Q}^H = (\mathbf{H}^H\mathbf{H})^{-1}$. Therefore, we have that when $2m \geq n$,
trace\{$\mathbf{Q}\mathbf{Q}^H$\} = trace\{$(\mathbf{H}^H\mathbf{H})^{-1}$\}.
{\color{black}
Then, it can be shown that
when $2m > n$,  $t^2=\mathcal{E} \{\text{trace} (\mathbf{Q}\mathbf{Q}^H)\}
=\mathcal{E} \{\text{trace}((\mathbf{H}^H\mathbf{H})^{-1})\}$.
Assume that  $\mathbf{H}_1$ and $\mathbf{H}_2$ are two  $m \times n$
matrices whose elements are i.i.d. complex Gaussian random variables with
zero mean and unit variance.
From \cite[Lemma 2.10]{tulino2004foundations}, it is easy to see that for $2m > n$,  $t^2=\mathcal{E} \{\text{trace} (\mathbf{Q}\mathbf{Q}^H)\}
=\mathcal{E} \{\text{trace}((\mathbf{H}^H\mathbf{H})^{-1})\}=\frac{1}{2\eta -1}$,
where $\eta=\frac{m}{n}$.
}

This completes the proof of  the theorem. \hspace{\fill}$\blacksquare$\newline

\vspace{-2em}

\section*{Appendix C: Proof of Theorem  \ref{Q33}}
{\color{black}
In this section, we characterize $t^2 =\mathcal{E} \{\text{trace} (\mathbf{Q}
\mathbf{B} \mathbf{Q}^H)\}$  when $2m < n$,
where  the $n \times n$ matrix $\mathbf{B}$
can be expressed as $\diag \left[ \mathbf{I}_m, \mathbf{O}_{(n-2m) \times (n-2m)}, \mathbf{I}_m  \right]$.}
From \eqref{GSVDFORM}, it can be shown that
\begin{eqnarray}
\left(\begin{array}{cc}
\mathbf{U}
& \mathbf{O}_{m \times m}
\\
 \mathbf{O}_{m \times m} & \mathbf{V}
\end{array}\right)
\left(\begin{array}{c}
\mathbf{H}_1
\\
 \mathbf{H}_2
\end{array}\right)\mathbf{Q}=
\left(\begin{array}{c}
\mathbf{\Sigma}_1
\\
\mathbf{\Sigma}_2
\end{array}\right).
\end{eqnarray}
Moreover, let us define $\mathbf{H}=\left(\begin{array}{c}
\mathbf{H}_1
\\
\mathbf{H}_2
\end{array}\right)$, $\mathbf{\Sigma}=\left(\begin{array}{c}
\mathbf{\Sigma}_1
\\
\mathbf{\Sigma}_2
\end{array}\right)$, $\mathbf{M}=\left(\begin{array}{cc}
\mathbf{U}
& \mathbf{O}_{m \times m}
\\
 \mathbf{O}_{m \times m} & \mathbf{V}
\end{array}\right)$. Then, {\color{black}it can be shown that} $\mathbf{H}=\mathbf{M}^H\mathbf{\Sigma} \mathbf{Q}^{-1}$.
Thus, we have that $(\mathbf{H}\mathbf{H}^H)^{-1}=(\mathbf{M}^H\mathbf{\Sigma} \mathbf{Q}^{-1}
\mathbf{Q}^{-H}\mathbf{\Sigma}^H \mathbf{M} )^{-1}$, and
\begin{eqnarray}
\label{qq1}
\text{trace}\{(\mathbf{H}\mathbf{H}^H)^{-1}\} &=& \text{trace}\{(\mathbf{M}^H\mathbf{\Sigma} \mathbf{Q}^{-1}
\mathbf{Q}^{-H}\mathbf{\Sigma}^H \mathbf{M} )^{-1}\}
\\ \nonumber
&=& \text{trace}\{\mathbf{M}^H(\mathbf{\Sigma} \mathbf{Q}^{-1}
\mathbf{Q}^{-H}\mathbf{\Sigma}^H  )^{-1} \mathbf{M}\}
\\ \nonumber
&=& \text{trace}\{\mathbf{M}\mathbf{M}^H(\mathbf{\Sigma} \mathbf{Q}^{-1}
\mathbf{Q}^{-H}\mathbf{\Sigma}^H  )^{-1} \}
\\ \nonumber
&=& \text{trace}\{ (\mathbf{\Sigma} \mathbf{Q}^{-1}
\mathbf{Q}^{-H}\mathbf{\Sigma}^H  )^{-1} \}.
\end{eqnarray}

From \eqref{mmn2}, {\color{black}it can be shown that}
\begin{eqnarray}
\label{appen11}
\mathbf{\Sigma}=\left(\begin{array}{ccc}
\mathbf{I}_m & \mathbf{O}_{m \times (n-2m)} & \mathbf{O}_{m \times m}
\\
\mathbf{O}_{m \times m} & \mathbf{O}_{m \times (n-2m)}  & \mathbf{I}_m
\end{array}\right),
\end{eqnarray}
and
$\mathbf{B}=\diag \left[ \mathbf{I}_m, \mathbf{O}_{(n-2m) \times (n-2m)}, \mathbf{I}_m  \right]=
\mathbf{\Sigma}^H\mathbf{\Sigma}$.
Then, we have
\begin{eqnarray}
\label{qq2}
\text{trace}\{\mathbf{Q}\mathbf{B}\mathbf{Q}^H\}
&=&\text{trace}\{\mathbf{Q}\mathbf{\Sigma}^H\mathbf{\Sigma}\mathbf{Q}^H\}
=\text{trace}\{\mathbf{\Sigma}\mathbf{Q}^H\mathbf{Q}\mathbf{\Sigma}^H\}.
\end{eqnarray}
Next, we show that $\mathbf{\Sigma}\mathbf{Q}^H\mathbf{Q}\mathbf{\Sigma}^H=(\mathbf{\Sigma} \mathbf{Q}^{-1}
\mathbf{Q}^{-H}\mathbf{\Sigma}^H  )^{-1}$.

Let us define $\mathbf{X}=\mathbf{U}_{11}\mathbf{H}_{12}'$.
From Appendix A, we know that when  $2m < n$, $\mathbf{Q}$ can be expressed as follows:
\begin{eqnarray}
\label{QStr}
\mathbf{Q}&=&\mathbf{Q}_1\mathbf{Q}_2 \mathbf{Q}_3
\\ \nonumber
&=& \mathbf{V}_2
\left(\begin{array}{cc}
\mathbf{I}_{n-m} & \mathbf{O}_{(n-m) \times m}\\
\mathbf{O}_{m \times (n-m)} & \diag(z_1, \cdots  z_m)^{-1}
\end{array}\right)
\left(\begin{array}{cc}
\mathbf{V}_{11} &\mathbf{O}_{(n-m) \times m}\\
\mathbf{O}_{m \times (n-m)}& \mathbf{I}_m
\end{array}\right)
\\ \nonumber&& \times
\left(\begin{array}{cc}
\diag(t_1, \cdots  t_m)^{-1} &\mathbf{O}_{m \times (n-m)}\\
\mathbf{O}_{(n-m) \times m}& \mathbf{I}_{n-m}
\end{array}\right)
\left(\begin{array}{cc}
\mathbf{I}_{n-m} & \left(\begin{array}{c}-\mathbf{X}
\\ \mathbf{O}_{(n-2m) \times m} \end{array}
\right)\\
\mathbf{O}_{m \times (n-m)}& \mathbf{I}_m
\end{array}\right).
\end{eqnarray}
{\color{black}Thus, using \eqref{QStr} and \eqref{appen11}}, it can be shown that
\begin{eqnarray}
\label{e1}
\mathbf{\Sigma}\mathbf{Q}^H\mathbf{Q}\mathbf{\Sigma}^H=
\left(\begin{array}{cc}
\diag(t_1, \cdots  t_m)^{-2} &   -\diag(t_1, \cdots  t_m)^{-2}  \mathbf{X}  \\
-\mathbf{X}^H\diag(t_1, \cdots  t_m)^{-2}& \mathbf{X}^H \diag(t_1, \cdots  t_m)^{-2} \mathbf{X}+\diag(z_1, \cdots  z_m)^{-2}
\end{array}\right).
\end{eqnarray}
Moreover, based on \eqref{QStr}, we can show that
\begin{eqnarray}
\label{apped3}
\mathbf{Q}^{-H}
&=& \mathbf{V}_2
\left(\begin{array}{cc}
\mathbf{I}_{n-m} & \mathbf{O}_{(n-m) \times m}\\
\mathbf{O}_{m \times (n-m)} & \diag(z_1, \cdots  z_m)
\end{array}\right)
\left(\begin{array}{cc}
\mathbf{V}_{11} &\mathbf{O}_{(n-m) \times m}\\
\mathbf{O}_{m \times (n-m)}& \mathbf{I}_m
\end{array}\right)
\\ \nonumber&&
\left(\begin{array}{cc}
\diag(t_1, \cdots  t_m) &\mathbf{O}_{m \times (n-m)}\\
\mathbf{O}_{(n-m) \times m}& \mathbf{I}_{n-m}
\end{array}\right)
\left(\begin{array}{cc}
\mathbf{I}_{n-m} & \mathbf{O}_{(n-m) \times m}\\
\left(\begin{array}{cc}
 \mathbf{X}^H & \mathbf{O}_{m \times (n-2m)}
\end{array}\right)
& \mathbf{I}_m
\end{array}\right).
\end{eqnarray}
{\color{black}Thus, based on \eqref{apped3} and \eqref{appen11}},
it can be obtained that
\begin{eqnarray}
\label{e2}
\mathbf{\Sigma} \mathbf{Q}^{-1}
\mathbf{Q}^{-H}\mathbf{\Sigma}^H =
\left(\begin{array}{cc}
\diag(t_1, \cdots  t_m)^2  +\mathbf{X}\diag(z_1, \cdots  z_m)^{2} \mathbf{X}^H  &  \mathbf{X}\diag(z_1, \cdots  z_m)^{2} \\
\diag(z_1, \cdots  z_m)^{2} \mathbf{X}^H& \diag(z_1, \cdots  z_m)^{2}
\end{array}\right).
\end{eqnarray}
Then, with \eqref{e1} and \eqref{e2}, {\color{black}it can be shown that} $ \mathbf{\Sigma}\mathbf{Q}^H\mathbf{Q}\mathbf{\Sigma}^H\mathbf{\Sigma} \mathbf{Q}^{-1}
\mathbf{Q}^{-H}\mathbf{\Sigma}^H= \mathbf{I}_{2m}$ and
\begin{eqnarray}
\label{qq3}
\mathbf{\Sigma}\mathbf{Q}^H\mathbf{Q}\mathbf{\Sigma}^H=(\mathbf{\Sigma} \mathbf{Q}^{-1}
\mathbf{Q}^{-H}\mathbf{\Sigma}^H  )^{-1}.
\end{eqnarray}

Finally, based on \eqref{qq1}, \eqref{qq2}, and \eqref{qq3}, {\color{black}it can be shown that} when $2m<n$,
$\text{trace}\{\mathbf{Q}\mathbf{B}\mathbf{Q}^H\}=\text{trace}\{\mathbf{\Sigma}\mathbf{Q}^H\mathbf{Q}\mathbf{\Sigma}^H\}
=\text{trace}\{ (\mathbf{\Sigma} \mathbf{Q}^{-1}
\mathbf{Q}^{-H}\mathbf{\Sigma}^H  )^{-1} \}=\text{trace}\{(\mathbf{H}\mathbf{H}^H)^{-1}\}$,
where $\mathbf{H}=\left(\begin{array}{c}
\mathbf{H}_1
\\
\mathbf{H}_2
\end{array}\right)$.
{\color{black}
Assume that  $\mathbf{H}_1$ and $\mathbf{H}_2$ are two  $m \times n$
matrices whose elements are i.i.d. complex Gaussian random variables with
zero mean and unit variance.
Using \cite[Lemma 2.10]{tulino2004foundations},
it is easy to see that for $2m < n$,  $t^2=\mathcal{E} \{\text{trace} (\mathbf{Q}\mathbf{B}\mathbf{Q}^H)\}
=\mathcal{E} \{\text{trace}((\mathbf{H}\mathbf{H}^H)^{-1})\}=\frac{1}{1/(2\eta) -1}$,
where $\eta=\frac{m}{n}$.}

This completes the proof of  the theorem. \hspace{\fill}$\blacksquare$\newline

\vspace{-2em}

\section*{Appendix D: Proof of Corollary  \ref{cor1}}
{\color{black}
When $m \geq n$, from \eqref{r1} and \eqref{r3},
the information rate of user 1 can be expressed as follows:
\begin{eqnarray}
R_{1i} =
\left\{ \begin{array}{ll}
 \log \left(  1+\frac{P w_i^2  l_2^2}{ t^2 d_1^\tau N_0 (1+w_i^2)} \right) &   w_i^2 > \frac{d_1^\tau}{d_2^\tau} \\
\log \left(  1+\frac{P w_i^2  l_1^2}{P w_i^2  l_2^2+ t^2 d_1^\tau N_0 (1+w_i^2)} \right) & w_i^2 \leq \frac{d_1^\tau}{d_2^\tau}
\end{array} \right.,
\end{eqnarray}
where $w_i^2$, $i \in \{1,\cdots,n\}$, is the
squared generalized singular value.
From Theorem \ref{Q}, {\color{black}it can be shown that} $t^2=\frac{1}{2\eta -1}$, where $\eta=\frac{m}{n}$.
Then,  {\color{black} user 1's  average information rate, $\bar{R}_{1i}^n$,} can be expressed as
follows:
\begin{eqnarray}
\bar{R}_{1i}^n= \frac{1}{n}\sum_{i=1}^{n} R_{1i}
&=& \int_0^{\infty} R_{1i} d  \mathbf{F}_{w_i}^n
\nonumber \\   &=& \int_{\frac{d_1^\tau}{d_2^\tau}}^{\infty} \log \left(  1+\frac{P x  l_2^2}{\frac{1}{2\eta -1} d_1^\tau N_0 (1+x)} \right) d  \mathbf{F}_{w_i}^n(x) \\&&
+\int_0^{\frac{d_1^\tau}{d_2^\tau}} \log \left(  1+\frac{P x  l_1^2}{P x  l_2^2+ \frac{1}{2\eta -1} d_1^\tau N_0 (1+x)} \right) d  \mathbf{F}_{w_i}^n(x),\nonumber
\end{eqnarray}
where $\mathbf{F}_{w_i}^n(x)$ is the e.d.f. of the squared generalized singular values $w_i^2$, $i \in \{1,\cdots,n\}$,
of the two users. {\color{black}
Moreover,  Theorem \ref{GSVDvalue} shows that}
when $m\geq n$, almost surely, $\mathbf{F}_{w_i}^n(x)$,
converges, as $m,n \to \infty$ with $\frac{m}{n}\to \eta$, to
a nonrandom c.d.f. $\mathbf{F}_{w_i}(x)$, whose
p.d.f. is $f_{\frac{1}{\eta}, \frac{1}{\eta}}(x)$.
Furthermore, it is easy to see that both
$\log \left(  1+\frac{P x  l_2^2}{\frac{1}{2\eta -1} d_1^\tau N_0 (1+x)} \right)$
and $\log \left(  1+\frac{P x  l_1^2}{P x  l_2^2+ \frac{1}{2\eta -1} d_1^\tau N_0 (1+x)} \right)$
are continuous bounded
functions in the domain $[0,\infty]$.
Then, from the bounded convergence theorem \cite[Theorem 6.3]{couillet2011random},
{\color{black}it can be shown that
as $m,n \to \infty$ with $\frac{m}{n}\to \eta$, $\bar{R}_{1i}^n$ converges to $\bar{R}_{1i}$, where
\begin{eqnarray}
\bar{R}_{1i}&=&\int_{\frac{d_1^\tau}{d_2^\tau}}^{\infty} \log \left(  1+\frac{P x  l_2^2}{\frac{1}{2\eta -1} d_1^\tau N_0 (1+x)} \right) d  \mathbf{F}_{w_i}(x) \nonumber \\&&
+\int_0^{\frac{d_1^\tau}{d_2^\tau}} \log \left(  1+\frac{P x  l_1^2}{P x  l_2^2+ \frac{1}{2\eta -1} d_1^\tau N_0 (1+x)} \right) d  \mathbf{F}_{w_i}(x).
\end{eqnarray}}
Finally, when $m \geq n$, $m,n \to \infty$ with $\frac{m}{n}\to \eta$,
the normalized  average rate  of user $1$,  $R_1$, can be expressed as
$\frac{1}{\eta}\bar{R}_{1i}$ and the expression for $\frac{1}{\eta}\bar{R}_{1i}$
is shown in Corollary \ref{cor1}.
The normalized average individual rate of user 2
can be obtained in a similar way.}
Thus, the corollary is proved.

This completes the proof of  the corollary. \hspace{\fill}$\blacksquare$\newline

\linespread{1}

\bibliographystyle{IEEEtran}
\bibliography{IEEEfull,cz}
\end{document}